\documentclass[%
reprint,
superscriptaddress,
amsmath,amssymb,
aps, 
prc
]{revtex4-1}
\RequirePackage{fancyhdr}
\RequirePackage{graphicx}
\usepackage{placeins}
\RequirePackage{adjustbox}
\usepackage{rotating}
\usepackage{multirow}
\usepackage{geometry}
\usepackage{color}
\usepackage{array}
\usepackage{float}
\usepackage{units} 
\RequirePackage{amssymb}
\RequirePackage{epstopdf}
\RequirePackage{amsmath, amsfonts}

\usepackage{graphicx}

\begin{document}

\title{Electric and magnetic dipole strength in $^{112,114,116,118,120,124}$Sn}%

\author{S.~Bassauer}\email{sbassauer@ikp.tu-darmstadt.de}\affiliation{\TUDarm}
\author{P.~von~Neumann-Cosel}\email{vnc@ikp.tu-darmstadt.de}\affiliation{\TUDarm}
\author{P.-G.~Reinhard}\affiliation{\Erlangen}
\author{A.~Tamii}\affiliation{\RCNP}
\author{S.~Adachi}\affiliation{\RCNP}
\author{C.~A.~Bertulani}\affiliation{\TAM}
\author{P.~Y.~Chan}\affiliation{\RCNP}
\author{A.~D'Alessio}\affiliation{\TUDarm}
\author{H.~Fujioka}\affiliation{\IOT}
\author{H.~Fujita}\affiliation{\RCNP}
\author{Y.~Fujita}\affiliation{\RCNP}
\author{G.~Gey}\affiliation{\RCNP}
\author{M.~Hilcker}\affiliation{\TUDarm}
\author{T.~H.~Hoang}\affiliation{\RCNP}
\author{A.~Inoue}\affiliation{\RCNP}
\author{J.~Isaak}\affiliation{\TUDarm}\affiliation{\RCNP}
\author{C.~Iwamoto}\affiliation{\RIKEN}
\author{T.~Klaus}\affiliation{\TUDarm}
\author{N.~Kobayashi}\affiliation{\RCNP}
\author{Y.~Maeda}\affiliation{\Miyazaki}
\author{M.~Matsuda}\affiliation{\Tohoku}
\author{N.~Nakatsuka}\affiliation{\TUDarm}
\author{S.~Noji}\affiliation{\NSCL}
\author{H.~J.~Ong}\affiliation{\IMP}\affiliation{\RCNP}
\author{I.~Ou}\affiliation{\Okayama}
\author{N.~Pietralla}\affiliation{\TUDarm}
\author{V.~Yu.~Ponomarev}\affiliation{\TUDarm}
\author{M.S.~Reen}\affiliation{\Akal}
\author{A.~Richter}\affiliation{\TUDarm}
\author{M.~Singer}\affiliation{\TUDarm}
\author{G.~Steinhilber}\affiliation{\TUDarm}
\author{T.~Sudo}\affiliation{\RCNP}
\author{Y.~Togano}\affiliation{\Rikkyo}
\author{M.~Tsumura}\affiliation{\Kyoto}
\author{Y.~Watanabe}\affiliation{\Tokyo}
\author{V.~Werner}\affiliation{\TUDarm}

\newcommand{\TUDarm}{Institut f\"ur Kernphysik, Technische Universit\"at Darmstadt, D-64289 Darmstadt, Germany}
\newcommand{\Erlangen}{Institut f\"ur Theoretische Physik II, Universit\"at Erlangen, D-91058 Erlangen, Germany}
\newcommand{\RCNP}{Research Center for Nuclear Physics, Osaka University, Ibaraki, Osaka 567-0047, Japan}
\newcommand{\TAM}{Department of Physics and Astronomy, Texas A\&M University-Commerce, Commerce, Texas 75429, USA}
\newcommand{\IOT}{Department of Physics, Tokyo Institute of Technology, Tokyo 152-8551, Japan}
\newcommand{\RIKEN}{RIKEN, Nishina Center for Accelerator-Based Science, 2-1 Hirosawa, 351-0198 Wako, Saitama, Japan}
\newcommand{\Miyazaki}{Department of Applied Physics, Miyazaki University, Miyazaki 889-2192, Japan}
\newcommand{\Tohoku}{Department of Communications Engineering, Graduate School of Engineering, Tohoku University, Aramaki Aza Aoba, Aoba-ku, Sendai 980-8579, Japan}
\newcommand{\NSCL}{National Superconducting Cyclotron Laboratory, Michigan State University, East Lansing, Michigan 48824, USA}
\newcommand{\IMP}{Institute of Modern Physics, Chinese Academy of Sciences, Lanzhou, 730000, China}
\newcommand{\Okayama}{Okayama University, Okayama 700-8530, Japan}
\newcommand{\Akal}{Department of Physics, Akal University, Talwandi Sabo, Bathinda Punjab-151 302, India}
\newcommand{\Rikkyo}{Department of Physics, Rikkyo University, Tokyo, Japan}
\newcommand{\Kyoto}{Department of Physics, Kyoto University, Kyoto 606-8502, Japan}
\newcommand{\Tokyo}{Department of Physics, University of Tokyo, Tokyo 113-8654, Japan}

\date{\today}%


\begin{abstract}
\begin{description}
\item[Background] 
Inelastic proton scattering at energies of a few hundred MeV and very forward angles including $0^\circ$ has been established as a tool for the study of electric and magnetic dipole strength distributions in nuclei.
The present work reports a systematic investigation of the chain of stable even-mass tin isotopes.
\item[Methods] 
Inelastic proton scattering experiments were performed at the Research Center for Nuclear Physics, Osaka, with a \unit[295]{MeV} beam covering laboratory angles $0^\circ - 6^\circ$ and excitation energies \unit[$6 - 22$]{MeV}.
Cross sections due to $E1$ and $M1$ excitations were extracted with a Multipole Decomposition Analysis (MDA) and then converted to reduced transition probabilities with the "virtual photon method" for $E1$ and the "unit cross section method" for $M1$ excitations, respectively.
Including a theory-aided correction for the high excitation energy region not covered experimentally, the electric dipole polarizability was determined from the $E1$ strength distributions. 
\item[Results]
Total photoabsorption cross sections derived from the $E1$ and $M1$ strength distributions show significant differences compared to those from previous $(\gamma,xn)$ experiments in the energy region of the IsoVector Giant Dipole Resonance (IVGDR).   
The widths of the IVGDR deduced from the present data with a Lorentz parameterization show an approximately constant value of about 4.5 MeV in contrast to the large variations between isotopes observed in previous work.
The IVGDR centroid energies are in good correspondence to expectations from systematics of their mass dependence.
Furthermore, a study of the dependence of the IVGDR energies on bulk matter properties is presented.
The $E1$ strengths below neutron threshold show fair agreement with results from $(\gamma,\gamma^\prime)$ experiments on $^{112,116,120,124}$Sn in the energy region between 6 and \unit[7]{MeV}, where also isoscalar $E1$ strength was found for $^{124}$Sn.
At higher excitation energies large differences are observed pointing to a different nature of the excited states with small ground state branching ratios.
The isovector spin-$M1$ strengths exhibit a broad distribution between 6 and \unit[12]{MeV} in all studied nuclei.
\item[Conclusions]
The present results contribute to the solution of a variety of nuclear structure problems including the systematics of the energy and width of the IVGDR, the structure of low-energy $E1$ strength in nuclei, new constraints to Energy Density Functionals (EDFs) aiming at a systematic description of the dipole polarizability across the nuclear chart, from which properties of the symmetry energy can be derived, and the systematics of the isovector spin-$M1$ strength in heavy nuclei.
\end{description}
\end{abstract}

\maketitle

\section{Introduction}

Inelastic proton scattering at energies of a few hundred MeV and very forward angles including $0^\circ$ has been established in recent years as a new spectroscopic tool for the investigation of electric and magnetic dipole strength distributions in nuclei \cite{vonNeumann-Cosel2019}.  
Although the $(p,p^\prime)$ reaction is rather non-selective in general exciting electric and magnetic modes alike, in the particular kinematics of very small momentum transfer a selective excitation of $E1$ and $M1$ strength is observed due to the following features: 
($i$) the incident beam is relativistic and Coulomb excitation dominates the cross sections \cite{Bertulani1988}, and
($ii$) the effective proton-nucleus interaction \cite{Love1981} is dominated by isovector spinflip transitions with orbital angular momentum transfer $\Delta L = 0$, i.e. the analog of Gamow-Teller (GT) transitions.

At present, such experiments at scattering angles very close to zero degrees can be performed at the Research Center for Nuclear Physics (RCNP), Japan \cite{Tamii2009} and at the iThemba Laboratory for Accelerator Based Sciences (iThemba LABS), South Africa \cite{Neveling2011}.
Dispersion matching between the beams and the magnetic spectrometers used to detect the scattered particles allows for high-resolution measurements of the order $\Delta E/ E = (1 - 2) \times 10^{-4}$.
Here, we report the results of a study of the stable tin isotopes $^{112,114,116,118,120,124}$Sn performed at RCNP. 
A decomposition of the dominant $E1$ and spin-$M1$ modes can be achieved either by an MDA of the cross sections  \cite{Poltoratska2012} or independently by the measurement of a combination of polarization transfer observables \cite{vonNeumann-Cosel2019}. 
Good agreement of both methods was demonstrated for reference cases \cite{Tamii2011,Hashimoto2015,Martin2017} indicating that the much simpler measurement of cross sections using an unpolarized beam and employ the MDA thereof is sufficient.

The results allow to address a variety of nuclear structure problems of current interest.
Low-energy electric dipole strength in nuclei with neutron excess, commonly termed Pygmy Dipole Resonance (PDR), is currently a subject of intense experimental and theoretical activities \cite{Savran2013,Bracco2019}. 
It occurs at energies well below the IVGDR and exhausts a considerable fraction (up to about 10\%) of the photoabsorption cross sections in nuclei with a large neutron-to-proton ratio \cite{Adrich2005,Klimkiewicz2007,Wieland2009,Rossi2013}. 
The properties of the mode are claimed to provide insight into the formation of a neutron skin
\cite{Klimkiewicz2007,Piekarewicz2006,Tsoneva2008,Piekarewicz2011,Reinhard2010} and the density dependence of the symmetry energy \cite{Klimkiewicz2007,Carbone2010,Fattoyev2012,Tsang2012}, although this is questioned \cite{Reinhard2010,Reinhard2013,Reinhard2014}.
Furthermore, dipole strength in the vicinity of the neutron threshold $S_n$ has an impact on neutron-capture rates in the astrophysical $r$-process \cite{Goriely2004,Litvinova2009,Daoutidis2012}.
A study of $^{120}$Sn revealed a dramatic difference of the low-energy isovector $E1$ response measured with the $(p,p^\prime)$ \cite{Krumbholz2015} and $(\gamma,\gamma^\prime)$ \cite{Oezel-Tashenov2014} reactions, respectively. 
The present work establishes this as a general phenomenon for the chain of stable even-mass tin isotopes and discusses implications for the structure of the PDR.  

Most of the information on photoabsoption cross sections in heavy nuclei stems from two methods, viz.\  $(\gamma,\gamma^\prime)$ \cite{Kneissl2006} and $(\gamma,xn)$ \cite{Berman1975} reactions.
Both rely on the measurement of the emission probability from the excited state and thus on knowledge of the branching ratio of the particular decay.
In contrast, the $(p,p^\prime)$ cross sections are directly related to the photoabsorption cross sections.
The experiments cover an excitation energy region from well below $S_n$ across the IVGDR, thus avoiding the difficulties of matching results from the two different experimental techniques, particularly pronounced near $S_n$.  
The IVGDR in stable tin isotopes was investigated in a series of $(\gamma,xn)$ experiments by different laboratories \cite{Berman1975,Fultz1969,Lepretre1974,Sorokin1974,Sorokin1975,Utsunomiya2009,Utsunomiya2011}. 
The present work sheds new light on the significant differences observed in the energy region of the IVGDR.

The energy region studied in the present experiments also covers the major part relevant to a determination of the nuclear electric dipole polarizability \cite{Bohigas1981}.
There is renewed interest into the polarizability because Energy Density Functional (EDF) theory \cite{Bender2003} predicts a correlation with the neutron skin thickness \cite{Brown2000} and leading parameters of a Taylor expansion of the symmetry energy around saturation density \cite{Reinhard2010,Piekarewicz2012,Roca-Maza2013}. 
This provides important constraints for the Equation of State (EoS) of neutron-rich matter, a major topic of current nuclear structure research \cite{Roca-Maza2018} important for an understanding of astrophysical events like core-collapse supernovae \cite{Yasin2020}, the formation of neutron stars~\cite{Oezel2016}, or neutron star mergers \cite{Abbott2017}. 
The polarizability in the chain of proton-magic tin nuclei is of particular interest because the underlying structure changes little between neutron shell closures $N = 50$ and 82.  
Two different driving agents for the evolution of the dipole polarizability are conceivable, viz.\ neutron excess and the general trend with mass number $A$ (i.e., the radius) both dependent on the symmetry energy.  
Accordingly, a variety of model calculations have been performed for this case attempting to explore this connection
\cite{Tsoneva2008,Piekarewicz2006,
Daoutidis2012,Tsoneva2004,
Vretenar2004,Terasaki2006,Litvinova2008,Lanza2009,Litvinova2010, Avdeenkov2011,Piekarewicz2012,Papakonstantinou2014,Piekarewicz2014,Ebata2014,Yuksel2019}.
Including a model-aided correction for the high-energy part of the excitation spectrum, the systematics of the dipole polarizability in the stable tin isotope chain are extracted from the present data.
A partial account of this work has been given in Ref.~\cite{Bassauer2020}.

Finally, the data provide new results on $M1$ strength in heavy nuclei.
The IsoVector Spin $M1$ (IVSM1) resonance is a fundamental excitation mode of nuclei \cite{Heyde2010} with relevance to diverse fields like the description of neutral-current neutrino interactions in supernovae \cite{Langanke2004,Langanke2008}, $\gamma$-strength functions utilized for physics of reactor design \cite{Chadwick2011} or the modeling of reaction cross sections in large-scale nucleosynthesis network calculations \cite{Loens2012}.
Since the mode is related to transitions between spin-orbit partners, it provides information on the evolution of single-particle properties leading to new shell closures in neutron-rich nuclei \cite{Otsuka2020}.
Furthermore, the IVSM1 resonance is the isospin-analog of the GT resonance \cite{Fujita2011} and thus provides insight into the long-standing problem of quenching of the GT strength \cite{Osterfeld1992,Ichimura2006}.
Data in heavy spherical nuclei are scarce, essentially limited so far to $^{90}$Zr \cite{Rusev2013} and $^{208}$Pb \cite{Laszewski1988,Birkhan2016}. 

This article is organized as follows.
Section \ref{sec2} yields information on the experiment and the data analysis including the techniques used to determine the unknown isotopic enrichment of some of the targets. 
Section \ref{sec3} provides details of the MDA and the resulting multipole decomposed cross section spectra.  
Section \ref{sec4} presents the conversion to photoabsoprtion cross sections and their comparison to previous work.
It also includes an analysis of the sensitivity of the IVGDR centroid energies to bulk parameters of nuclear matter. 
The relevance of the new results on $E1$ and spin-$M1$ strength distributions to the various nuclear structure problems discussed above (PDR, polarizability, IVSM1 resonance) is discussed in Section~\ref{sec5}.  
A summary and an outlook on future work is given in the concluding Section~\ref{sec6}.

\section{Experiment and data analysis}
\label{sec2}

\subsection{Experimental details}

The inelastic proton scattering experiments were performed in 2015 and 2017 at RCNP.
In 2015, $^{112,116,124}$Sn and with lesser statistics $^{118,120}$Sn were measured. 
In the second experimental campaign in 2017, $^{112,116,124}$Sn were measured again to improve statistics. 
Additionally, data on $^{114,118,120,122}$Sn were taken. The measurements used the Grand Raiden spectrometer \cite{Fujiwara1999}. 
Data were taken at central spectrometer angles of 0$^\circ$, 2.5$^\circ$ and 4.5$^\circ$. 
Typical beam currents were \unit[$2-20$]{nA}, depending on the spectrometer angle. 
The unpolarized incident proton beam had an energy of \unit[295]{MeV}.
Dispersion matching and background optimization at $0^\circ$ were performed following the procedures described in Ref.~\cite{Tamii2009}. 
The energy resolutions achieved varied between 30 and \unit[40]{keV} (Full Width at Half Maximum, FWHM). 
At the end of both experimental campaigns sieve slit measurements were made with a thick $^{58}$Ni target to obtain precise angle calibrations. 
Additionally, elastic scattering data for all investigated tin isotopes were taken in the first experimental campaign.

\begin{table}
	\centering
	\caption[Targets used during the experiments.]
			{Targets used during the experiments.
			Given are the areal density $\rho$x, the enrichment and the purpose of the corresponding target.}
	\begin{tabular}{cccc}
 		\hline
 		\hline
 		\noalign{\vskip 0.5mm}
 		Target & $\rho$x & Enrichment & Purpose \\
 			& (mg/cm$^2$) & (\%) &  \\
 		\hline
 		\noalign{\vskip 0.5mm}
 		$^{112}$Sn & 3.38 & 90.2(1.4) & main target \\
 		$^{112}$Sn & 10.3 & 95.1($<$1) & calibration \\
 		$^{114}$Sn & 7.51 & 87.1($<$1) & main target \\
 		$^{116}$Sn & 4.98 & 95.5($<$1) & main target \\
 		$^{116}$Sn & 4.65 & 97.8($<$1) & main target \\
 		$^{118}$Sn & 4.50 & 86(7) & main target \\
 		$^{120}$Sn & 6.50 & 98.4($<$1) & consistency check \\
 		$^{124}$Sn & 5.00 & 97.0($<$1) & main target \\
 		$^{124}$Sn & 4.67 & 97.4($<$1) & main target \\
 		$^{197}$Au & 1.68 & 100 & beam tuning \\
 		$^{26}$Mg  & 1.16 & unknown& energy calibration \\
 		$^{58}$Ni  & 100.1& unknown & sieve slit \\
 		$^{12}$C   & 1.01 & 98.9 & energy calibration \\
 		C$_2$H$_4$ & 2.30 & $-$	& beam tuning \\
    	\hline
    	\hline
	\end{tabular}
	\label{tab:targets}
\end{table}
A summary of the used targets is given in Tab. \ref{tab:targets}. 
All tin targets were highly enriched self-supporting metallic foils with areal densities between 3.4 and \unit[7.5]{mg/cm$^2$}. 
The uncertainties of the target enrichment are quoted by the supplier to be better than \unit[1]{\%}.
However, in some cases the enrichment was unknown. 
In the case of $^{112}$Sn, a second thicker target with a known enrichment was used to determine the enrichment of the thinner target, but could only be measured in achromatic mode with corresponding reduced resolution due to its limited extension and the high areal density of \unit[10.3]{mg/cm$^2$}.
After folding to obtain the same energy resolution, the abundance was determined to 90.2(1.4)\% by normalization of the two spectra. 
The enrichment of $^{118}$Sn was estimated from the systematics of the IVGDR after conversion to  photoabsorption cross sections. 
A Lorentzian fit shows a smooth dependence of the centroid energy and width on the mass number as discussed  below in Sec.~\ref{subsec:ivgdr}. 
By interpolating these integrated values, the enrichment for $^{118}$Sn was determined to 86(7)\%.
A presumably enriched $^{122}$Sn target was also measured.
However, the IVGDR properties deviated significantly from the systematics, and low-energy spectra taken at larger scattering angles showed a broad bump in the energy region \unit[$2-3$]{MeV} instead of the expected excitation of known $2^+$ states in $^{122}$Sn.
Both findings suggested a natural isotopic composition. 
Thus, the spectra were discarded from further analysis.
Data for $^{120}$Sn were taken to check the consistency with a previous experiment of the same type \cite{Hashimoto2015,Krumbholz2015}. 

After each measurement of a main target for one hour, a short run with $^{12}$C was performed for energy calibration and to account for possible energy shifts of the beam. 
Further data for the energy calibration were taken using $^{26}$Mg and polyethylene (C$_2$H$_4$) targets.
The areal densities of the tin targets quoted by the manufacturer were remeasured and the corresponding uncertainties were determined to be around \unit[5]{\%}. 

\subsection{Particle Identification}

 A distinction of protons from other ejectiles can be achieved by investigating the deposited energy in the plastic scintillator trigger detectors. Furthermore, the particles can also be discriminated by their time of flight (ToF). 
 The ToF information is obtained from the trigger signal generated by one of the scintillators and from the radio frequency of the Azimutally Varying Field (AVF) cyclotron. 
 To improve the particle identification, the ToF information was linearly corrected to make it independent of the horizontal position $x_{fp}$ in the focal plane and of the horizontal scattering angle $\theta_{fp}$. 

\begin{figure}
	\centering
	\includegraphics[width=\columnwidth]{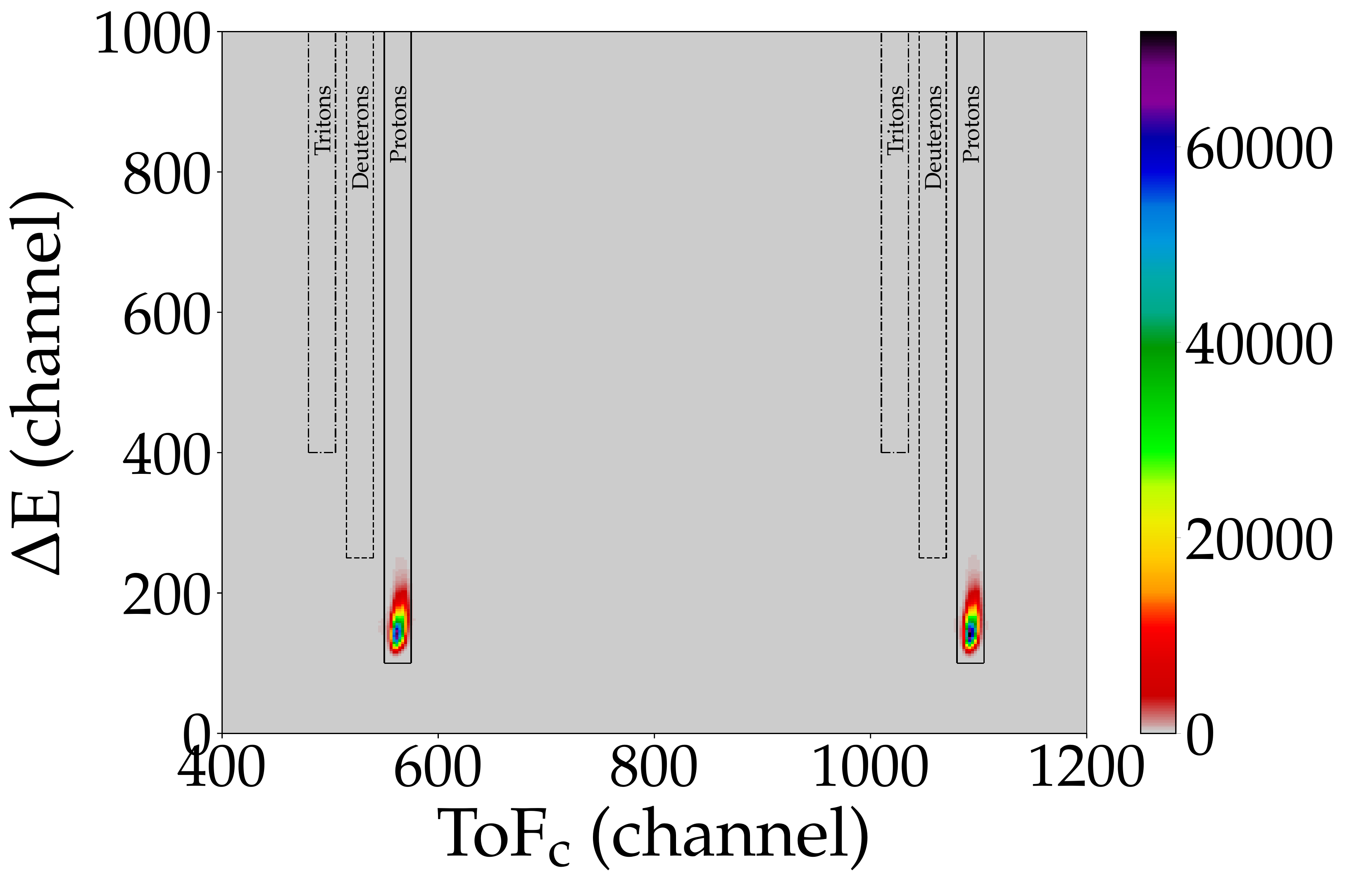}
	\caption{Particle identification via the correlation of energy loss and corrected time of flight (ToF$_{\rm c}$). 
	Two beam bunches are shown. 
	The time difference between the two bunches corresponds to a beam pulse period of about \unit[60]{ns}.}
	\label{fig:tof}
\end{figure}
In Fig.~\ref{fig:tof}, the energy loss $\Delta E$ in the plastic scintillator is plotted against the corrected time of flight $\mathrm{ToF_c}$.
The proton scattering events, framed by a two-dimensional rectangular gate, can be clearly identified. 
Predicted regions for deuteron and triton events (e.g.\ from $(p,d)$ and $(p,t)$ reactions) are also indicated. 
However, neither deuterons nor tritons were observed in this experiment.

\subsection{Angle calibration}

To obtain a precise angle calibration, sieve slit measurements were performed with a thick $^{58}$Ni target under a spectrometer angle of $16^\circ$. 
In order to investigate the dependence of the scattering angle on the horizontal position at the focal plane, five different magnetic field settings were measured covering the entire momentum acceptance of the spectrometer. 
Additionally, the vertical beam position at the target was changed by 0 and \unit[$\pm$1]{mm} relative to the centre, so that three measurements per magnetic field setting were realised.
The reconstruction of the horizontal and vertical scattering angles with a multidimensional least-squares fitting analysis followed the approach described in Ref.~\cite{Tamii2009}.

\subsection{Energy calibration}

\begin{figure}
	\centering
	\includegraphics[width=\columnwidth]{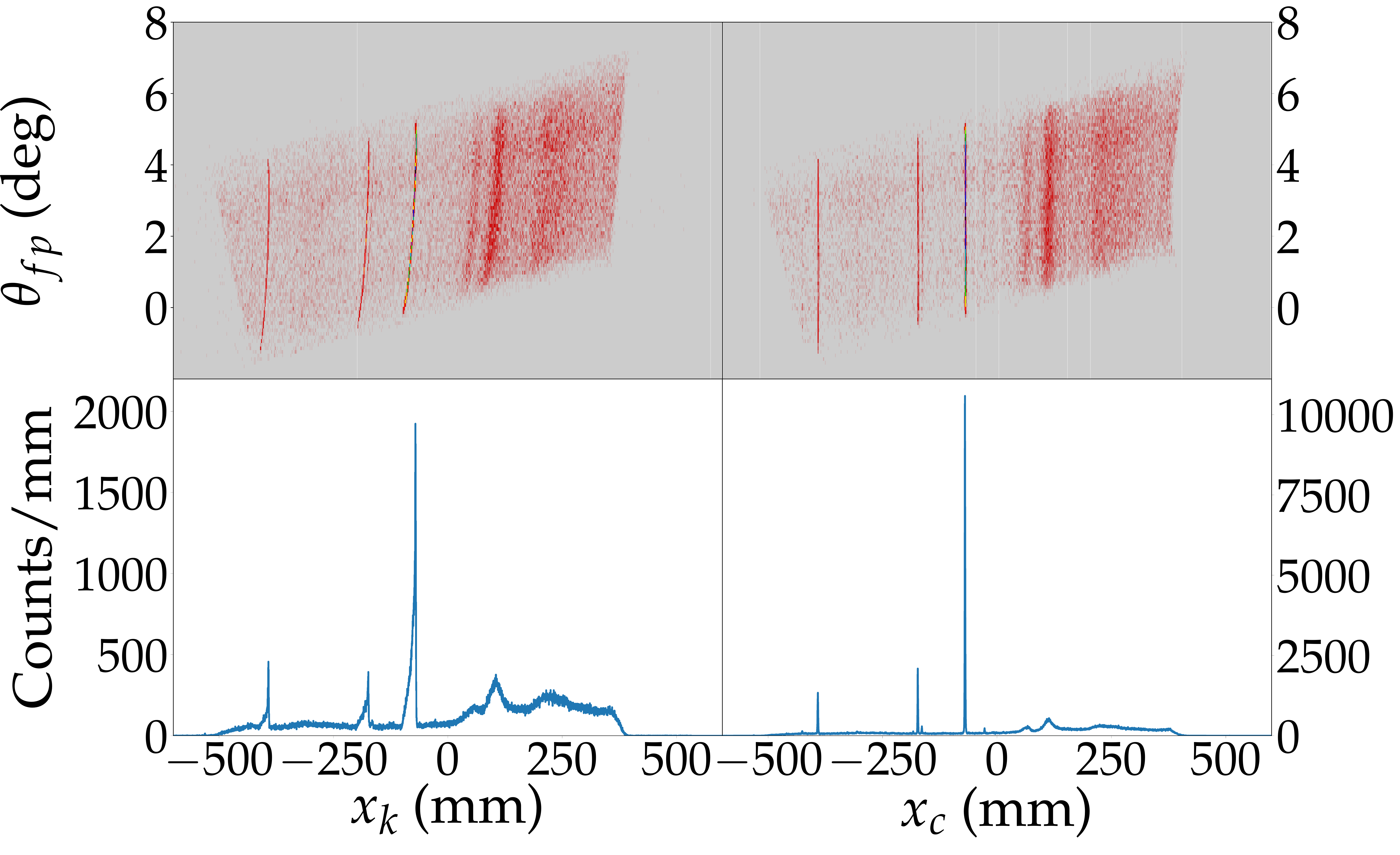}
	\caption{Focal plane spectra of the $^{12}$C$(p,p^\prime)$ reaction before (left) and after (right) the aberration correction described in the text.}
	\label{fig:energycalibration}
\end{figure}

In order to achieve an optimum energy resolution, the  correlation between the horizontal position $x_{k}$ and the horizontal scattering angle at the focal plain $\theta_{fp}$ due to the ion-optical properties of the Grand Raiden Spectrometer \cite{Fujiwara1999} needs to be corrected. 
On the left side of Fig.~\ref{fig:energycalibration} data for $^{12}$C are shown in the $\theta_{fp}-x_k$ plane as well as their projection on the abscissa. 
One can clearly see the most prominently excited states in $^{12}$C at 7.6, 12.7 and \unit[15.1]{MeV} (from left to right).
The curvature visible in the two-dimensional correlation leads to an asymmetric line shape in the projected energy spectrum distorting the resolution.
A two-dimensional least-squares fit was performed to account for this.
The result is depicted on the right side of Fig.~\ref{fig:energycalibration}, where $x_c$ denotes the corrected position on the focal plane. 
The resolution is improved considerably from about \unit[180]{keV} to \unit[30]{keV} (FWHM).

Excitation energies were determined from a second-order polynomial fit of the focal plane position of well-known transitions in the calibration spectra determined assuming Gaussian line shapes.  
Using these calibration functions, the reference energies of the known transitions could be reconstructed to \unit[$\pm$4]{keV} in the excitation energy region from \unit[5]{MeV} to \unit[18]{MeV}. 
The average energy resolution achieved was \unit[40]{keV} (FWHM)  in the first and \unit[30]{keV} (FWHM) in the second campaign, respectively.

\subsection{Background subtraction}

The main contribution of the experimental background at very forward angles stems from multiple scattering of incident protons in the target material. 
Scattering off the beam pipes or slits also contributes occasionally, especially during the $0^\circ$ measurements. 
Due to the statistical nature of multiple scattering, a flat distribution of background events is expected on the focal plane in non-dispersive direction $y_{fp}$, while true events are concentrated around $y_{fp}=0$.

\begin{figure}[b]
	\centering
	\includegraphics[width=\columnwidth]{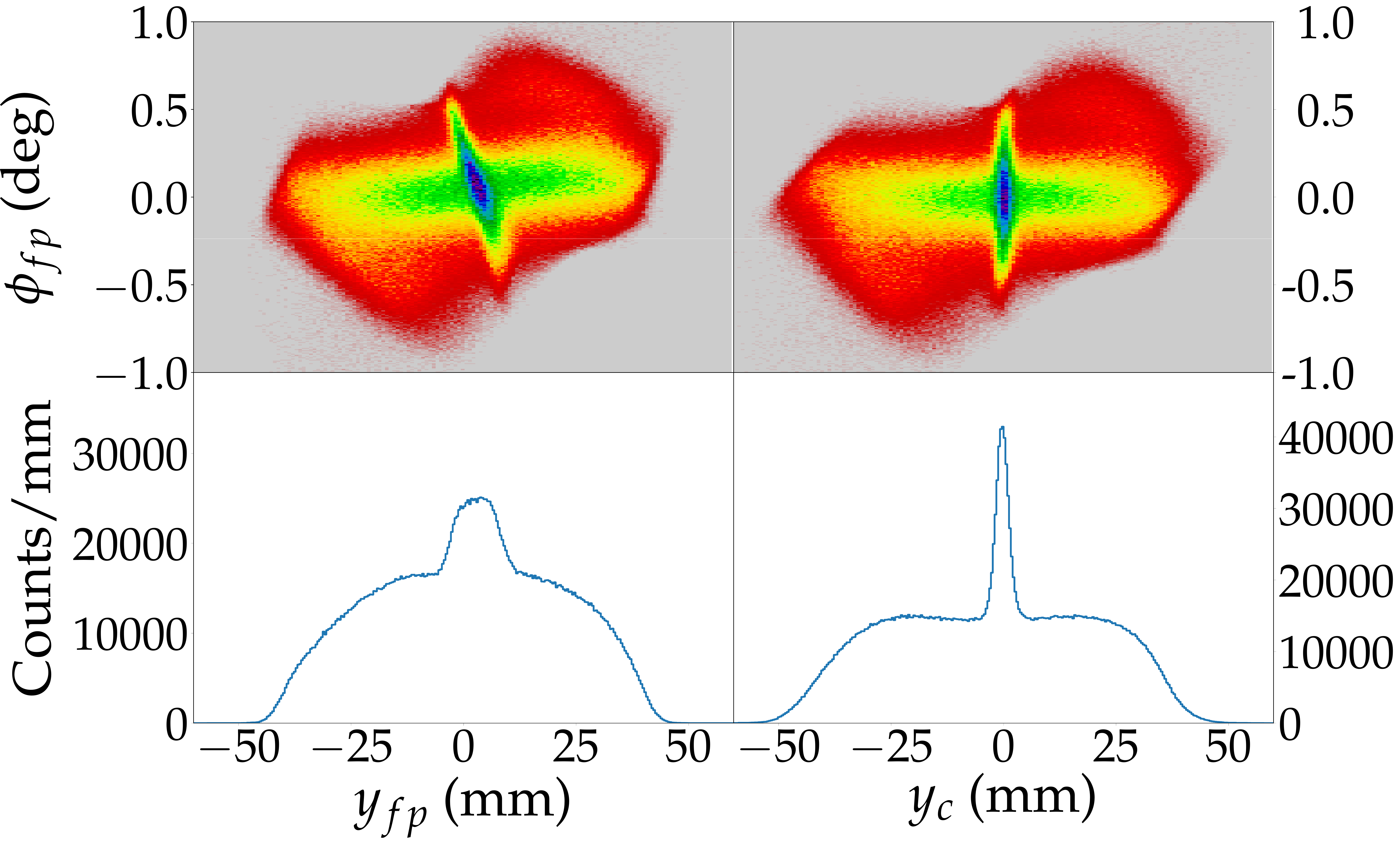}
	\caption{Correlation of the position ($y_{fp}$)  in the non-dispersive direction and the vertical angle ($\phi_{fp}$) before (left) and after (right) the restoration of the focusing condition at the focal plane. 
	The vertical angle was corrected in such a way that the background events are distributed symmetrically around $y_{c}=0$.}
	\label{fig:underfocuscorrection}
\end{figure}
However, the operation of the Grand Raiden Spectrometer in the so-called underfocus mode~\cite{Fujita2001} necessary to improve the resolution of the vertical angle leads to a dependence of $y_{fp}$ on the vertical scattering angle $\phi_{fp}$ as illustrated on the left side of Fig.~\ref{fig:underfocuscorrection}. 
Hence, before the background can be determined, a correction of $y_{fp}$ needs to be carried out to restore the focusing condition at the focal plane. 
This can be achieved with a multidimensional least-squares fit as a function of the position and scattering angles plus a correction for the vertical position, see Ref.~\cite{Tamii2009}.

The effect of the correction can be seen in Fig.~\ref{fig:underfocuscorrection}, where the correlation  between $y_{fp}$ and the vertical angle $\phi_{fp}$ before and after the restoration of the focusing condition at the focal plane is compared.
The corrected spectrum exhibits the expected flat background distribution.

\begin{figure}
	\centering
	\includegraphics[width=\columnwidth]{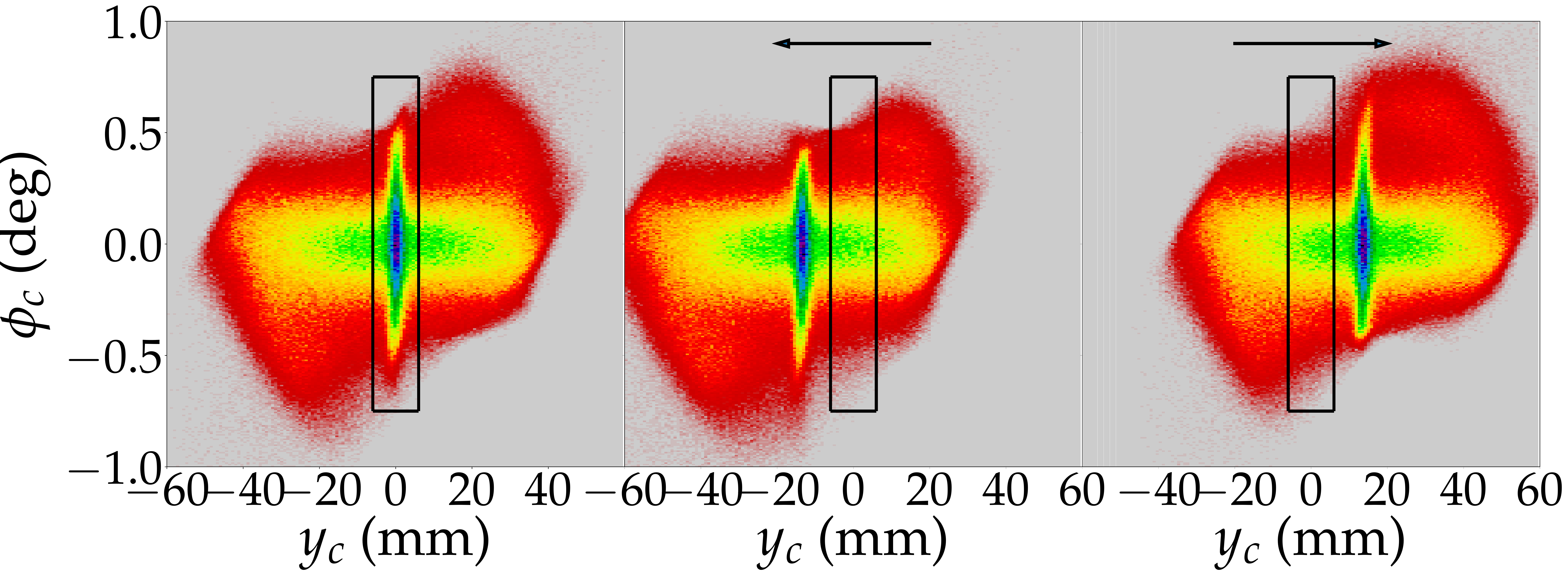}
	\caption{Background subtraction procedure using the correlation of the position $y_c$ in non-dispersive direction and the vertical angle $\phi_{fp}$. 
	The two-dimensional gate including true and background events is indicted by the black rectangle. 
	To determine the background, the data were shifted by a constant value to the left and right in $y_c$ direction. 
	}
	\label{fig:backgroundsubtraction}
\end{figure}

After the correction, the background was determined in the following way. 
Three data sets were generated as illustrated in Fig.~\ref{fig:backgroundsubtraction}. 
In the first data set a gate was set to contain the
true plus the background events. 
The second and the third data sets were analyzed in
exactly the same way as the first one, except that the data were shifted along the $y_c$ axis by a constant value. 
After the shift, the gate only contained background events. 

\begin{figure}
	\centering
	\includegraphics[width=0.8\columnwidth]{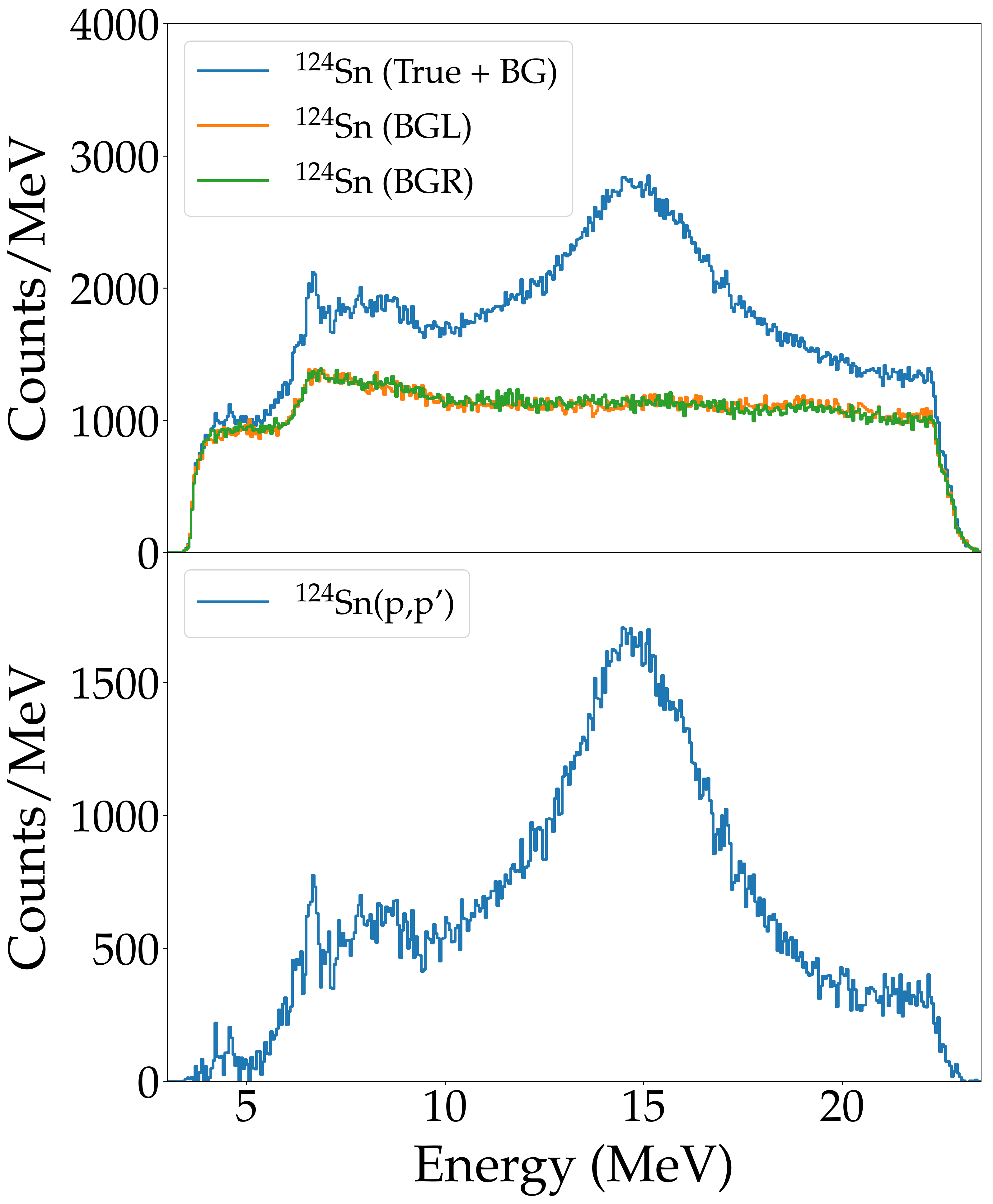}
	\caption{Top: Excitation energy spectra of $^{124}$Sn measured at $0^\circ$ corresponding to the three data sets from Fig.~\ref{fig:backgroundsubtraction} true-plus-background (blue), background from shift to the left (orange), and to the right (green). 
	Bottom: Background-subtracted spectrum.}
	\label{fig:spectrabgnobg}
\end{figure}

The background events from the shifted data sets were then averaged and finally subtracted from the first data set.
The energy spectra corresponding to the three data sets are displayed in the upper part of Fig.~\ref{fig:spectrabgnobg} for the example of $^{124}$Sn measured at $0^\circ$, where the blue histogram corresponds to true-plus-background and the orange and green histograms to the background spectra after the shift to left and right in Fig.~\ref{fig:backgroundsubtraction}, respectively.
As expected, the pure background spectra from the two shifted data sets are identical within statistical uncertainties.
The lower part of Fig.~\ref{fig:spectrabgnobg} present a background-free spectrum after subtraction of the averaged contribution from the orange and green spectra.

\subsection{Cross sections and uncertainties}

Absolute double differential cross sections were determined from the experimental parameters: collected charge, target properties, drift chamber efficiency, spectrometer solid angle and data acquisition dead time.
For the procedures to extract these quantities from the raw data see Ref.~\cite{Bassauer2019}.

The total cross section uncertainties were calculated taking statistical and systematic uncertainties in quadrature.
Major contributions to the systematic uncertainties originate from the solid angle determination (\unit[$4-5$]{\%}),
target thickness (\unit[5]{\%}) and charge collection (\unit[3]{\%}), whereas all other contributions to the systematic
uncertainty are \unit[$<$1]{\%}.

\section{Multipole decomposition}
\label{sec3}

An MDA of the measured spectra has been performed to extract $E1$ and $M1$ cross sections. 
The MDA is well established in the analysis of giant resonances \cite{Harakeh2001} and has been applied for charge-exchange reactions \cite{Wakasa1996,Wakasa1997} aiming at the extraction of the Gamow-Teller strength, but also in inelastic alpha scattering \cite{Bonin1984,Li2010,Itoh2013} used to study isoscalar giant resonances. 
It also serves as a reliable tool in the analysis of inelastic proton scattering data~\cite{Poltoratska2012, Krumbholz2015, Martin2017}.

\subsection{Experimental spectra}

\begin{figure}
	\centering
	\includegraphics[width=\columnwidth]{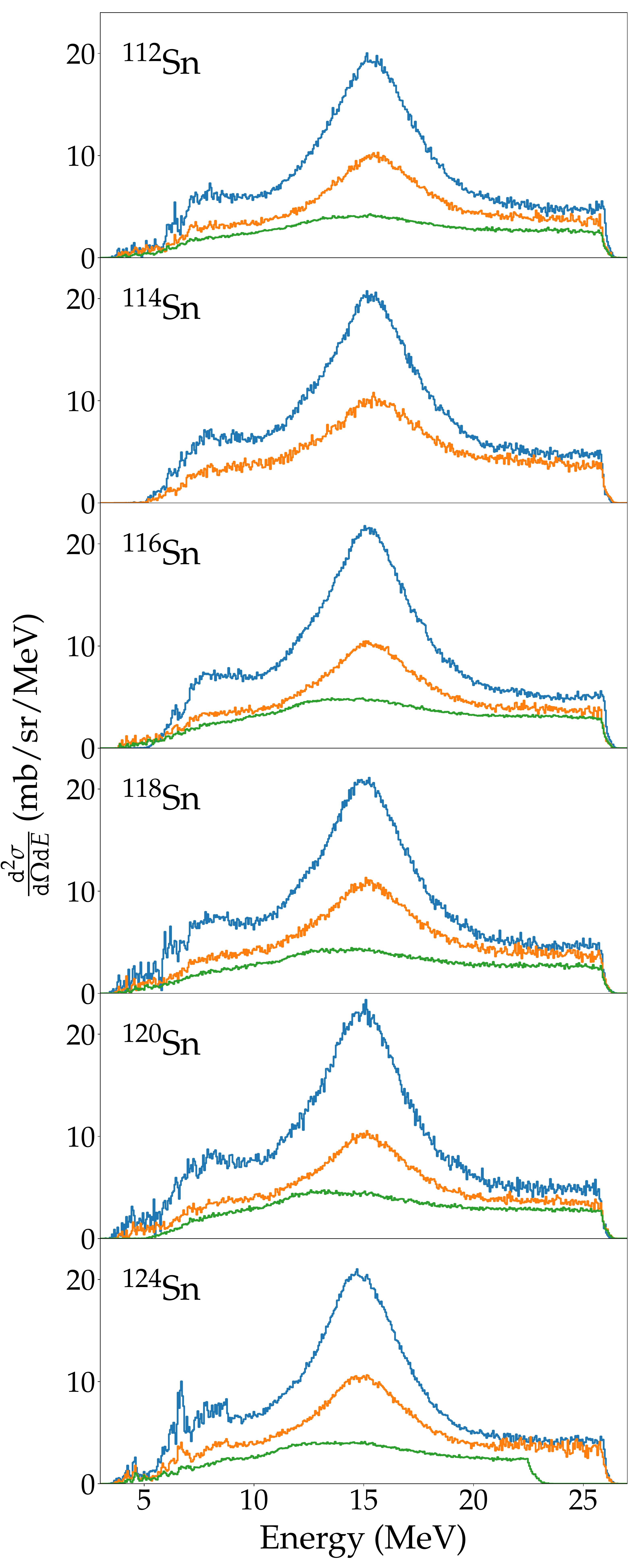}
	\caption{Double differential cross sections of the $^{112,114,116,118,120,124}$Sn$(p,p^\prime)$ reactions at \unit[$E_0 = 295$]{MeV} for spectrometer angles $\Theta=0^\circ$ (blue), $\Theta=2.5^\circ$ (orange) and $\Theta=4.5^\circ$ (green).}
	\label{fig:ddcs}
\end{figure}

The double differential cross sections extracted as described in the previous Section are summarized in Fig.~\ref{fig:ddcs}.
Data at $4.5^\circ$ are missing for $^{114}$Sn due to the lack of beam time. 
For $^{124}$Sn, data at $4.5^\circ$ were only taken in the first experimental campaign. 
Therefore, the excitation energy spectrum extends only up to about \unit[23.5]{MeV}, due to different magnetic field settings. 
In all isotopes, the GDR can be clearly identified around \unit[15]{MeV}. 
In the PDR region between 6 and \unit[10]{MeV}, a structure can be seen becoming gradually more pronounced for heavier isotopes culminating in $^{124}$Sn, where even distinct peaks are formed. 
The typical decrease of the cross section with increasing angle due to dominant Coulomb excitation is apparent both in the PDR and GDR energy regions. 

\subsection{Theoretical input}

\begin{figure}
	\centering
	\includegraphics[width=\columnwidth]{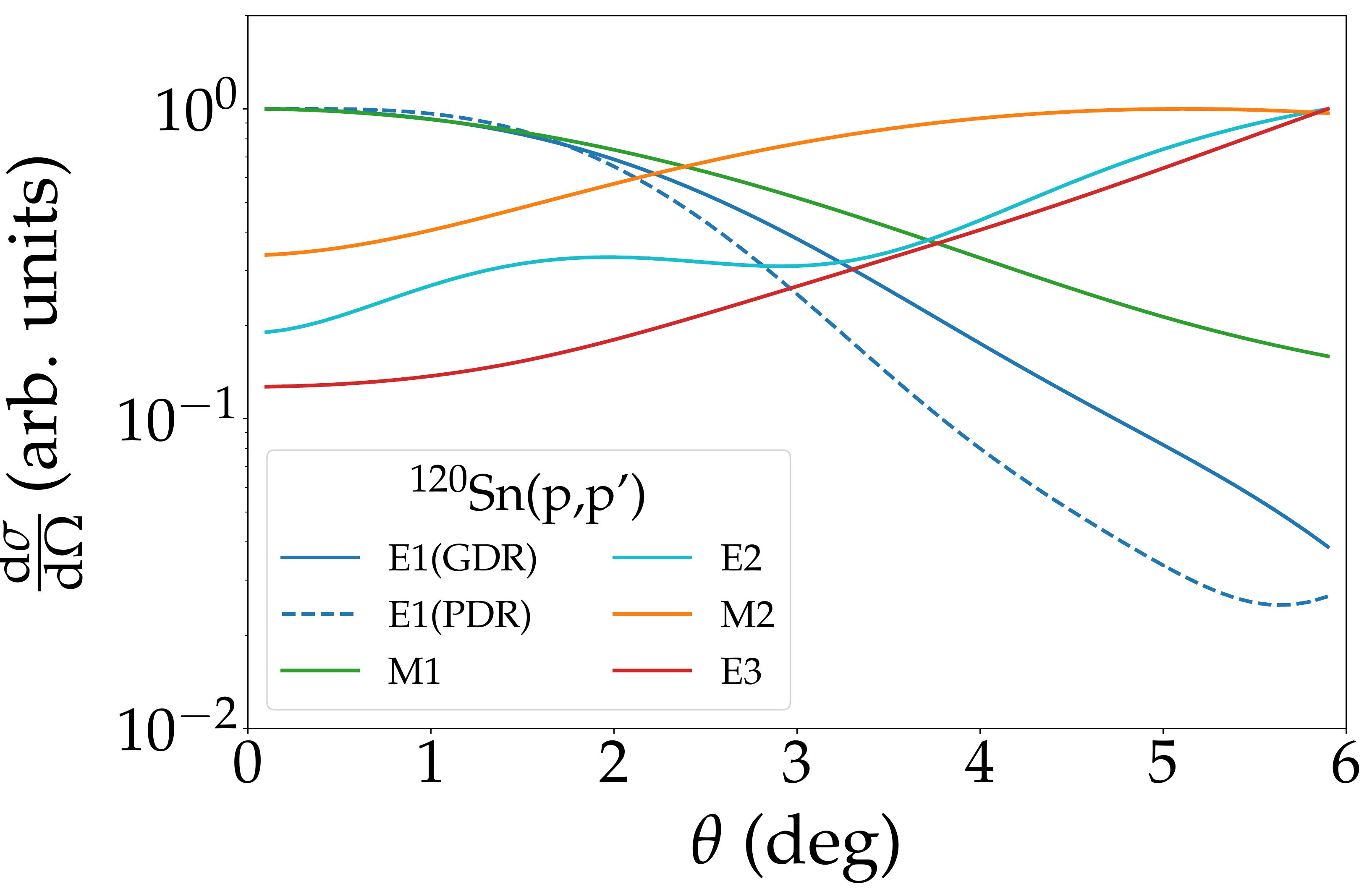}
	\caption{Angular distributions of different multipolarities calculated with the code DWBA07 and QPM transition densities for the $^{120}$Sn$(p,p^\prime)$ reaction in the angular range $0^\circ - 6^\circ$. 
	The maxima of the curves are normalised to unity.}
	\label{fig:120SnCurves}
\end{figure}

Theoretical angular distributions of the differential cross sections for different multipolarities were calculated using the code DWBA07 \cite{dwba07}.
Transition amplitudes and single-particle wave functions obtained from Quasiparticle Phonon Model (QPM) calculations of the type described in Refs.~\cite{Poltoratska2012,Krumbholz2015} were used as input. 
The parameterization of Love and Franey~\cite{Love1981} was employed to describe the effective nucleon-nucleon interaction.
An example of the angular distributions of different multipolarities is shown in Fig.~\ref{fig:120SnCurves} for the case of $^{120}$Sn.

The shapes suggest that $E1$ and $M1$ contributions are dominant under small angles, whereas higher multipolarities, such as $E2$, $M2$, and $E3$, are only relevant for larger angles in the experimentally studied range. 
The theoretical curves of Fig.~\ref{fig:120SnCurves} were also used for the MDA of all other tin isotopes, since the underlying structure for the tin isotopes studied in this work is very similar and the angular distributions of collective modes show a weak dependence on mass number. 
They were, however, corrected for the slightly different recoil term depending on the isotope masses and convoluted with the experimental angular resolution.

\subsection{Subtraction of the ISGMR and ISGQR}
\label{SubGMRGQR}

\begin{figure}
	\centering
	\includegraphics[width=\columnwidth]{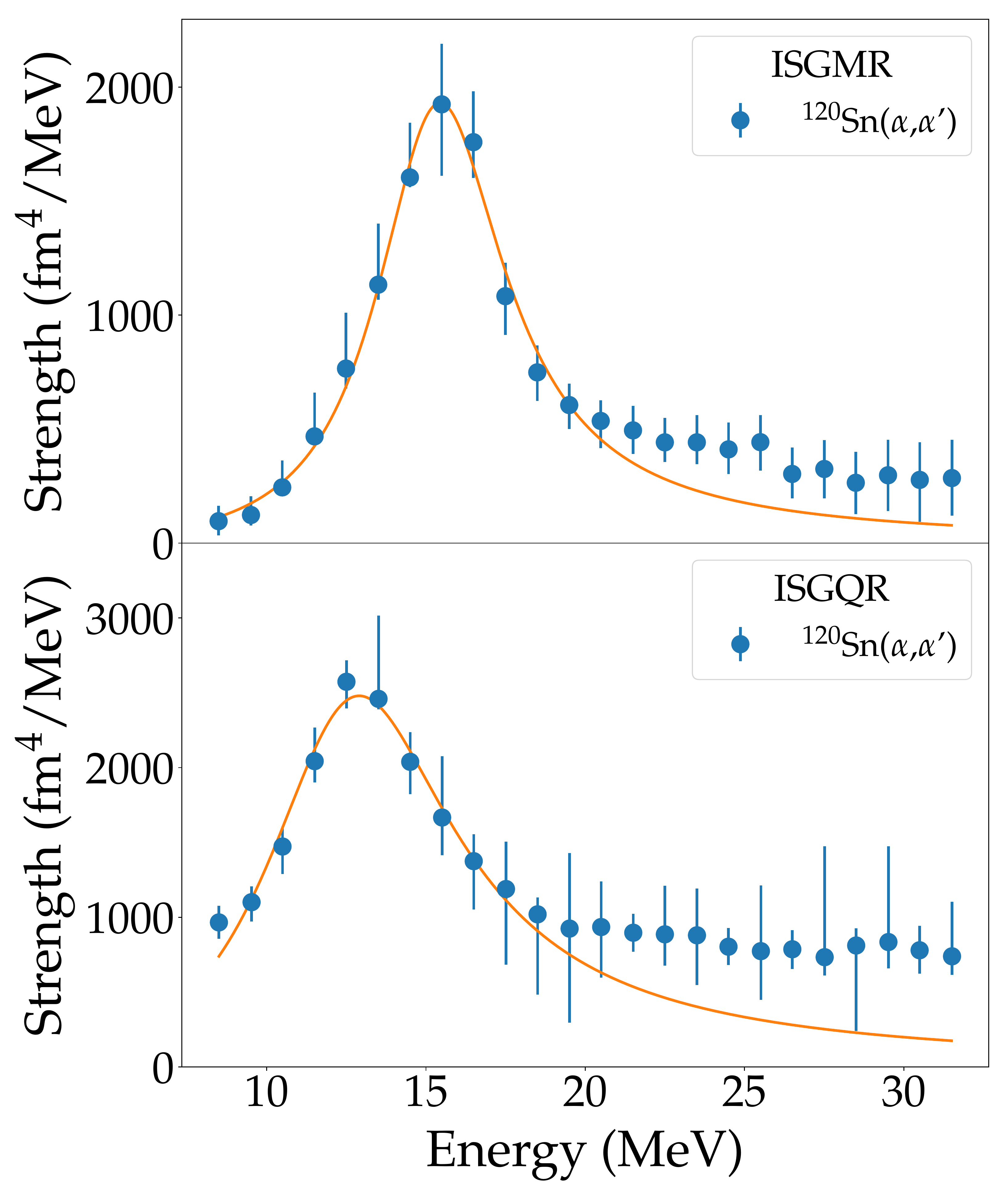}
	\caption{Strength distribution of the ISGMR (top) and ISGQR (bottom) in $^{120}$Sn from $\alpha$ scattering experiments~\cite{Li2010}. 
	Lorentzian fits in the resonance region are shown as orange curves.}
	\label{fig:GMRGQRstrength}
\end{figure}

Since the number of data points available is limited to 15 (5 per spectrometer angle), the number of multipolarities in the MDA must also be limited to avoid ambiguities.
One particular problem is the excitation of the IsoScalar Giant Monopole Resonance (ISGMR), which has an angular distribution similar to the $E1$ and $M1$ cases.
The contributions of the ISGMR and the Isoscalar Giant Quadrupole Resonance (ISGQR) were subtracted prior to performing the MDA. 
Experimental information on these modes is available from ($\alpha$,$\alpha^\prime$) experiments~\cite{Li2010} for all tin isotopes in question. 
The corresponding strength distributions for the example of $^{120}$Sn are presented in Fig.~\ref{fig:GMRGQRstrength}.
The orange curves are Lorentzian fits in the resonance region.
We note that only the Lorentzians were used for the subtraction procedure described below, because at higher excitation energies contributions from continuum scattering are included.

\begin{figure}
	\centering
	\includegraphics[width=\columnwidth]{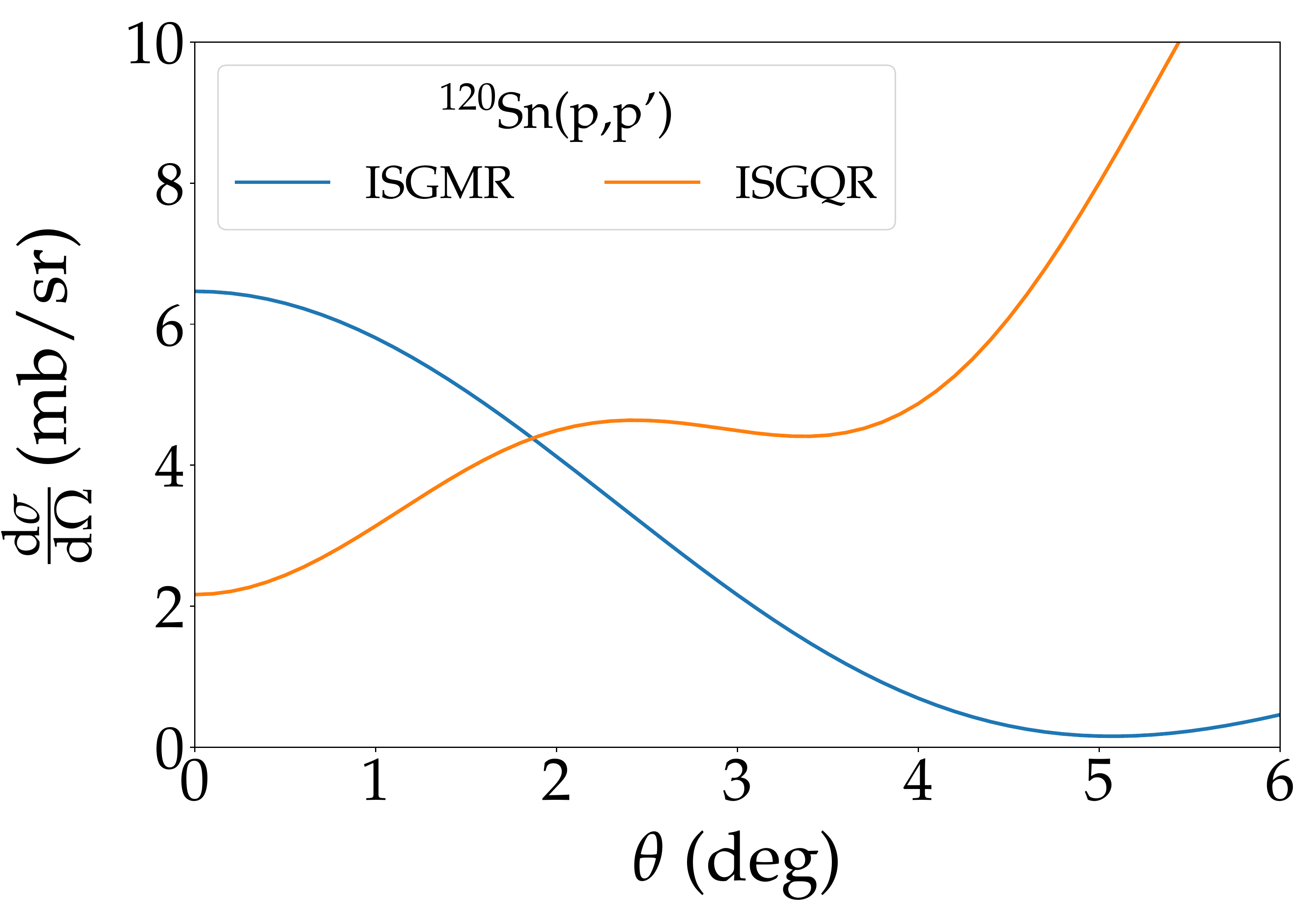}
	\caption{Theoretical $(p,p^\prime)$ cross sections of the ISGMR and ISGQR in $^{120}$Sn calculated with the DWBA07 code. }
	\label{fig:GMRGQRDWBA}
\end{figure}

The contribution of the ISGMR and ISGQR to the proton scattering cross sections can be estimated with the following approach \cite{Donaldson2018}
\begin{equation}
\label{eqn:gmrgqr}
	\frac{\mathrm{d}\sigma}{\mathrm{d}\Omega}(\theta, E_x)=
	\frac{\mathrm{d}\sigma}{\mathrm{d}\Omega}(\theta)_{DWBA}
	\frac{IS(E\lambda)(E_x)_{\rm exp}}{IS(E\lambda)_{\rm th}},
\end{equation}
where $IS(E\lambda)(E_x)_{\rm exp}$ are the isoscalar strength distributions from $\alpha$ scattering and $IS(E\lambda)_{\rm th}$ the theoretical strength from QPM calculations with $\lambda=0$ for ISGMR and $\lambda=2$ for ISGQR, respectively. 
Equation~(\ref{eqn:gmrgqr}) makes use of the fact that
inelastic proton scattering at energies of a few hundred MeV is a direct process and one can assume proportionality between the strength and the cross sections. 
The theoretical strength distributions were calculated within the QPM and the strongest $E0$ and $E2$ transitions were then utilized to determine cross sections using the DWBA07 code.
The theoretical cross sections shown in Fig.~\ref{fig:GMRGQRDWBA} correspond to about 50\% and 100\% of the ISGMR and ISGQR EWSR, respectively. 

\begin{figure}
	\centering
	\includegraphics[width=\columnwidth]{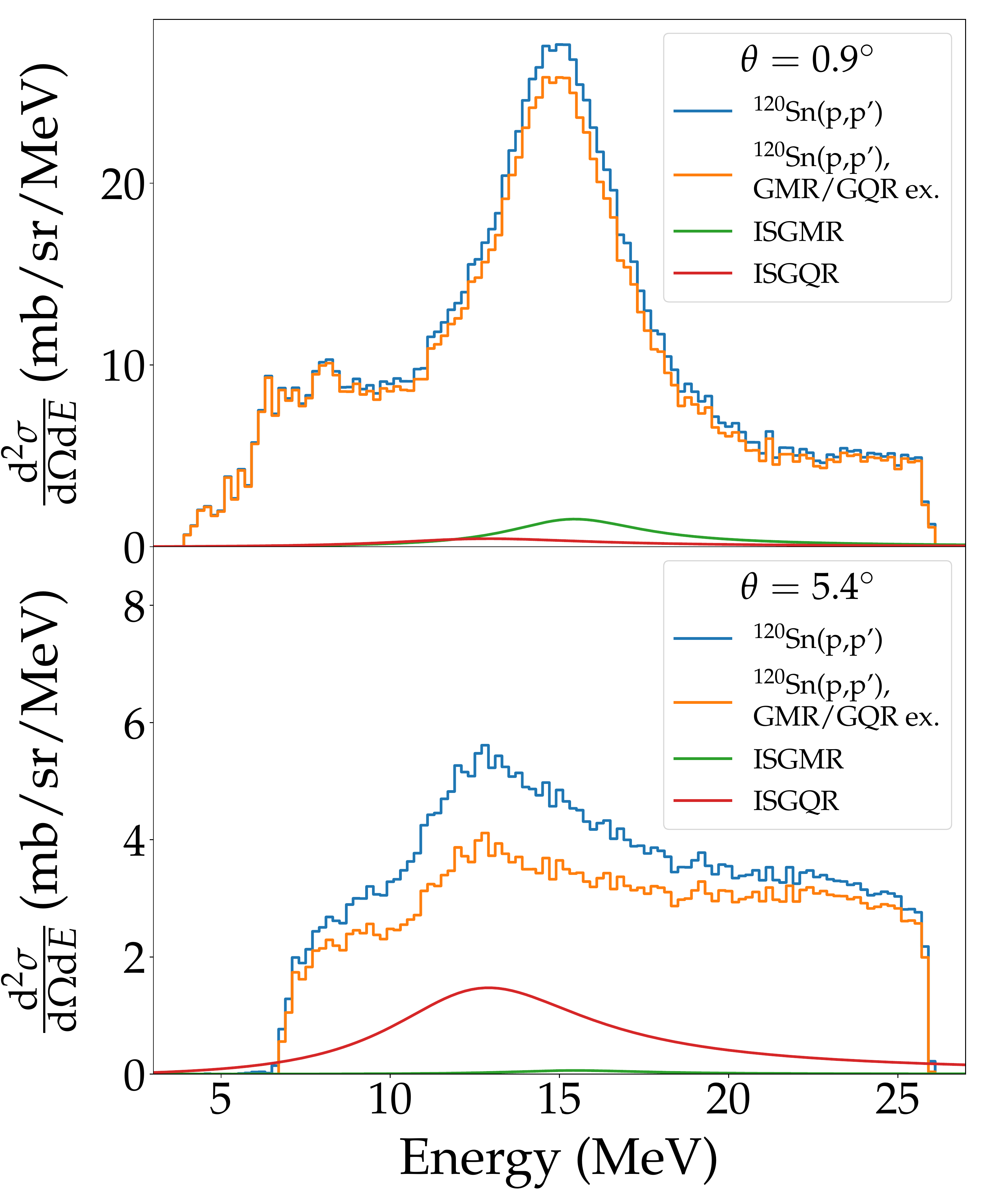}
	\caption{Spectra of the $^{120}$Sn$(p,p^\prime)$ reaction before (blue) and after (orange) subtraction of the ISGMR (green) and ISGQR (red) contributions for two different angles.
	}
	\label{fig:GMRGQRcorr}
\end{figure}

Finally, in Fig.~\ref{fig:GMRGQRcorr} the estimated contributions of the ISGMR and ISGQR to the experimental spectra of $^{120}$Sn are presented for two angles at $0.9^\circ$ and $5.4^\circ$. 
They are rather small for the very forward angle. 
The monopole contribution is more important but never exceeds \unit[5]{\%}. 
For larger angles however, a considerable contribution from the ISGQR is found reaching \unit[25]{\%} at the maximum, while the ISGMR contribution is negligible. 
After the subtraction of the ISGMR and ISGQR contributions a bump around \unit[13]{MeV} can still be seen for the $5.4^\circ$ data. 
This suggests that the absolute cross section of the ISGQR might be underestimated, though possible contributions from higher multipolarities, such as M2 and E3, were not considered yet which could possibly explain the remaining bump. 
Since all higher multipoles show a similar angular distribution at larger angles  (above $3^\circ$ in the present case) in Fig.~\ref{fig:120SnCurves}, possible remaining ISGQR contributions are accounted for by allowing one representative multipolarity $\lambda > 1$ in the MDA \cite{Poltoratska2012}.
The possible impact on the decomposition of E1 and M1 cross sections is nevertheless small, because the latter is mainly determined at the most forward angles. 

\subsection{Continuum background}
\label{QFSBackground}

\begin{figure}
	\centering
	\includegraphics[width=\columnwidth]{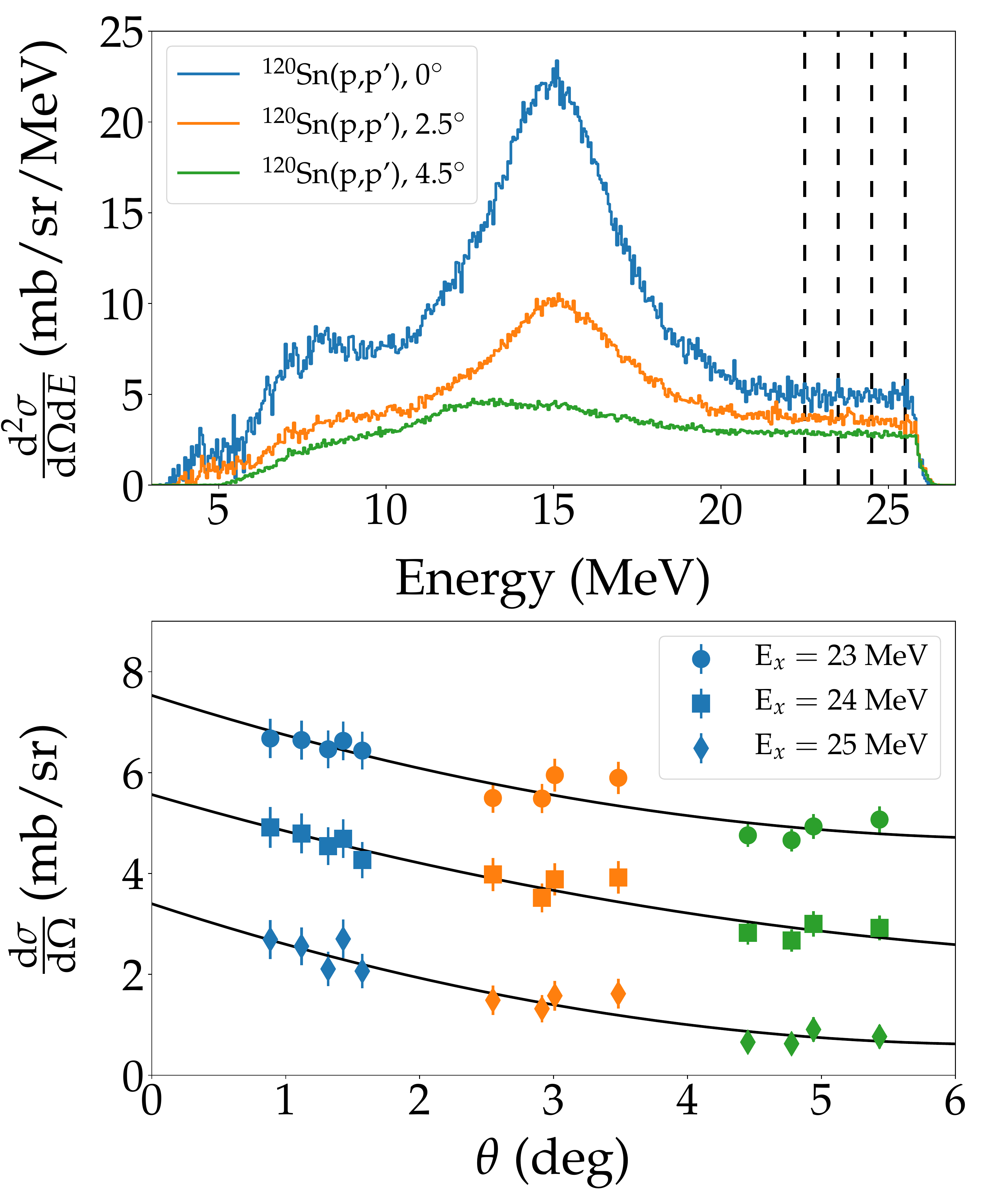}
	\caption{Top: Excitation energy spectra of $^{120}$Sn and excitation energy bins (vertical dashed lines) used to determine a parameterization of the angular dependence of the continuum background. 
	Bottom: Corresponding angular distributions for different energy bins together with the fit of Eq.~(\ref{eq:qfs}).
	For better visibility, they are shifted relative to each other by \unit[2]{mb/sr}.}
	\label{fig:PhenBG}
\end{figure}

Besides the excitation of electric and magnetic resonances the spectra also contain a continuum part, which dominates the spectra at energies above the IVGDR and needs to be taken into account for in the MDA.
It is believed to result mainly from quasifree scattering (QFS), although other contributions are not excluded.  
The QFS process occurs only at energies above the particle thresholds.  
In Ref.~\cite{Poltoratska2011}, a phenomenological parameterization was determined for the $(p,p^\prime)$ reaction on $^{208}$Pb and a similar approach was used in this work based on the $^{120}$Sn spectra. 
The nucleus $^{120}$Sn was chosen because it is the heaviest measured nucleus with data available for all three measured angles in the high excitation energy region, where possible contributions from the high-energy tail of the IVGDR are negligible. 
The data were analyzed in \unit[1]{MeV} bins to reduce statistical fluctuations, and angular distributions in the energy region between 22.5 and \unit[25.5]{MeV} were extracted.
The angular distributions were then fitted with polynomial functions of second order. 
Since these were identical within error bars for all bins in the selected energy region, an average polynomial function 
\begin{equation}
	\frac{\mathrm{d}\sigma}{\mathrm{d}\Omega}(\theta)_{\rm BG} = 5.7(3)-1.0(2) \theta+0.09(3) \theta^2.
	\label{eq:qfs}
\end{equation}
was determined for the background component.
The upper part of Fig.~\ref{fig:PhenBG} displays the $^{120}$Sn data used. 
The energy bins chosen for the angular distributions are indicated by the vertical dashed lines. 
In the lower part, the angular distributions for the three energy bins are shown together with the fit given in Eq.~(\ref{eq:qfs}).
For better visibility, they are shifted relative to each other by a constant (\unit[2]{mb/sr}).
Note that only 4 angular gates were applied to the data taken at finite spectrometer angles because of limited statistics.
Equation (\ref{eq:qfs}) describes all data well and also scales well with the results of a similar analysis of the $^{208}$Pb data \cite{Tamii2011} if the mass ratio is taken into account.

\subsection{Results}

\begin{figure*}
	\centering
	\includegraphics[width=0.8\textwidth]{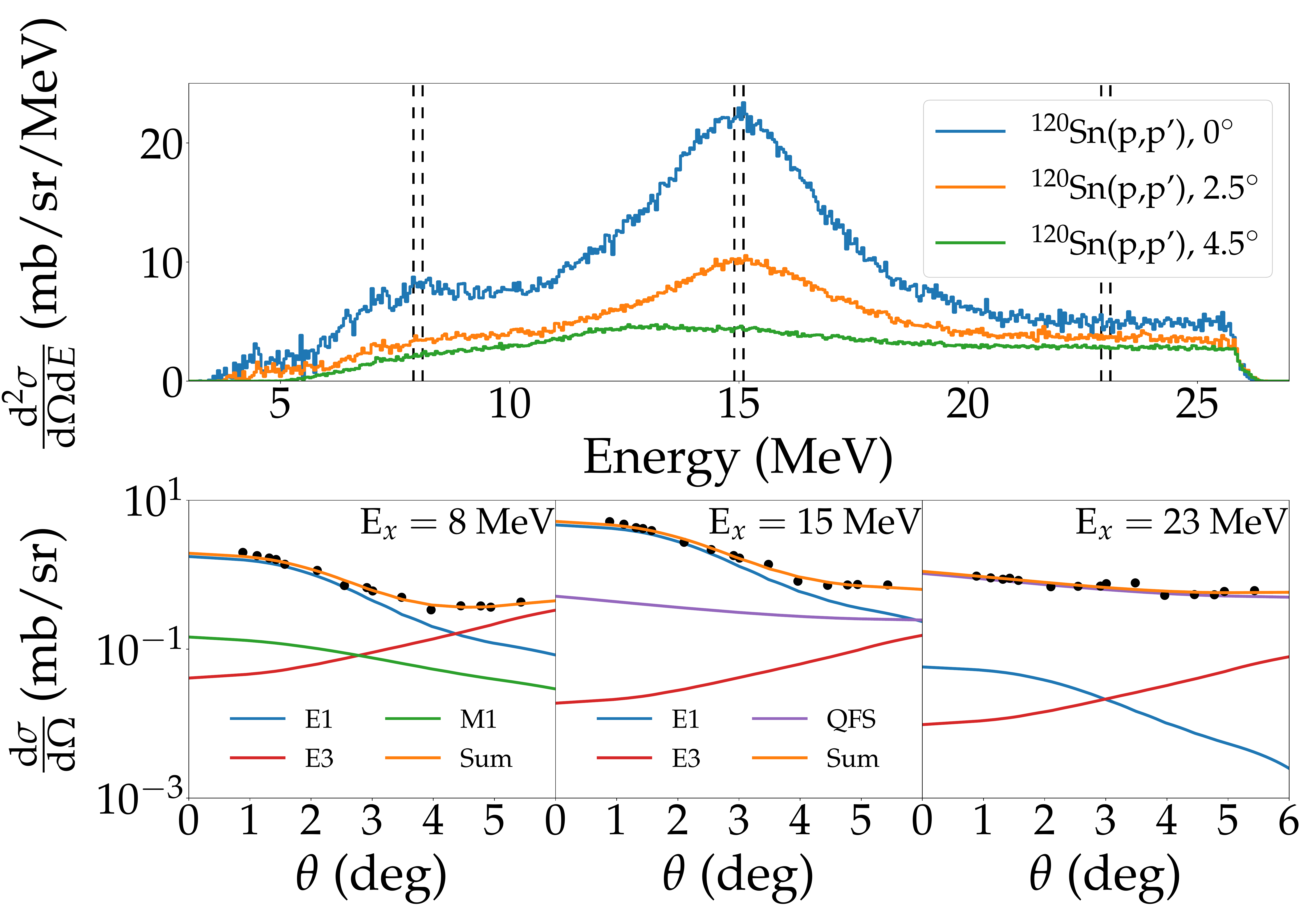}
	\caption{Typical results of the MDA for the example of $^{120}$Sn and three different energy bins at 8, 15 and \unit[23]{MeV}.
	Top: Spectra and energy bins indicated by the vertical dashed lines.
	Bottom: Experimental angular distributions and results of Eq.~(\ref{eq:chi2red}) for different multipoles and their sum.}
	\label{fig:MDAEx}
\end{figure*}

For the MDA all spectra were rebinned to \unit[200]{keV} and the ISGMR and ISGQR contributions were subtracted as described in Sec.~\ref{SubGMRGQR}. 
Experimental angular distributions of the differential cross sections for each bin were then determined and the data were fitted by means of a least-squares method with linear combinations of the theoretically predicted angular distributions of the differential cross sections via

\begin{equation}
 \sum_{i} 
 \left(\frac{\textnormal{d}\sigma}{\textnormal{d}\Omega}(\theta_i,E_{\rm x})_{\textnormal{exp}} - \frac{\textnormal{d}\sigma}{\textnormal{d}\Omega}(\theta_i,E_{\rm x})_{\textnormal{th}}\right)^2 = {\rm min},
\label{eq:fit}
\end{equation}
with
\begin{align}
\frac{\textnormal{d}\sigma}{\textnormal{d}\Omega}(\theta,E_{\rm x})_{\textnormal{th}} =&
\sum_{\pi \lambda} a_{\pi \lambda}  \frac{\textnormal{d}\sigma}{\textnormal{d}\Omega}(\theta,E_{\rm x},\pi\lambda)_{\textnormal{DWBA}} \nonumber \\
& +b\frac{\textnormal{d}\sigma}{\textnormal{d}\Omega}(\theta)_{\textnormal{BG}},
\label{eq:theo}
\end{align}
where $a_{\pi \lambda}$ and $b$ are fit parameters. 
The fits were performed using the following criteria and boundary conditions: \\
-- For each data set measured at the spectrometer angle $\theta = (0^\circ, 2.5^\circ, 4.5^\circ)$, five data points per angle and energy bin were generated by applying gates to the vertical and horizontal angles respectively, so that in total 15 data points between $0.9^\circ$ and $5.4^\circ$ were available for the MDA.\\
-- In total six $E1$ transitions (three in the PDR region and three in the GDR region) with the largest $B(E1)$ values in the QPM calculations were used, since the corresponding angular distributions show sensitivity to the Coulomb-nuclear interference. \\
-- Two $M1$ transitions with the largest $B(M1)$ values in the QPM calculations were used.\\
-- The $E3$ transition was used as a substitute for possible higher multipole contributions. \\
-- Equation (\ref{eq:qfs}) was used for the continuum background.\\
-- All parameters  $a^{\pi\lambda}$ and $b$ had to be positive.

The least-squares fitting procedure was carried out including all possible combinations of the theoretical angular distributions satisfying the above criteria. 
For each combination the $\chi^2$ and the reduced $\chi^2_{red}=\chi^2/(p-n)$ values were calculated with $p$ the number of experimental data points and $n$ the number of fit parameters. 
Using $\omega=1/\chi^2_{red}$ as a weighting parameter, mean cross sections for each contribution were finally determined
\begin{equation}
	\left\langle\frac{\mathrm{d}\sigma}{\mathrm{d}\Omega}(\theta, E_x)^{\pi\lambda}\right\rangle = \frac{\sum_i\omega_i \frac{\mathrm{d}\sigma}{\mathrm{d}\Omega}(\theta, E_x)_i^{\pi\lambda}}{\sum_i\omega_i}.
	\label{eq:chi2red}
\end{equation}
The corresponding uncertainty was obtained from the weighted variance
\begin{equation}
	\sigma^2 = \frac{\sum_i\omega_i \left(\frac{\mathrm{d}\sigma}{\mathrm{d}\Omega}(\theta, E_x)_i^{\pi\lambda}-\left\langle\frac{\mathrm{d}\sigma}{\mathrm{d}\Omega}(\theta, E_x)^{\pi\lambda}\right\rangle\right)^2}{\sum_i\omega_i}.
	\label{eq:chi2rederror}
\end{equation}

In Fig.~\ref{fig:MDAEx} a typical result of the MDA is displayed for the example of $^{120}$Sn and three different energy bins at 8, 15 and \unit[23]{MeV}.
The upper part shows the $^{120}$Sn spectra and the energy bins indicated by vertical dashed lines.
In the lower part the corresponding experimental angular distributions and the results of Eq.~(\ref{eq:chi2red}) for different multipoles and their sums are given.
$E1$ cross sections are largest in the PDR region (\unit[8]{MeV}), but the $M1$ contribution at angles close to $0^\circ$ is non-negligible.
At larger angles some higher-multipole component is needed to account for the data.
The energy bin near the maximum of the IVGDR (\unit[15]{MeV}) exhibits the expected dominance of $E1$ cross sections at forward angles.
The only other relevant contribution is the continuum background.
Finally, at the high excitation energy (\unit[23]{MeV}) all multipole contributions are at least more than an order of magnitude weaker than the continuum cross sections. 

\begin{figure}
   \centering
	\includegraphics[width=0.89\columnwidth]{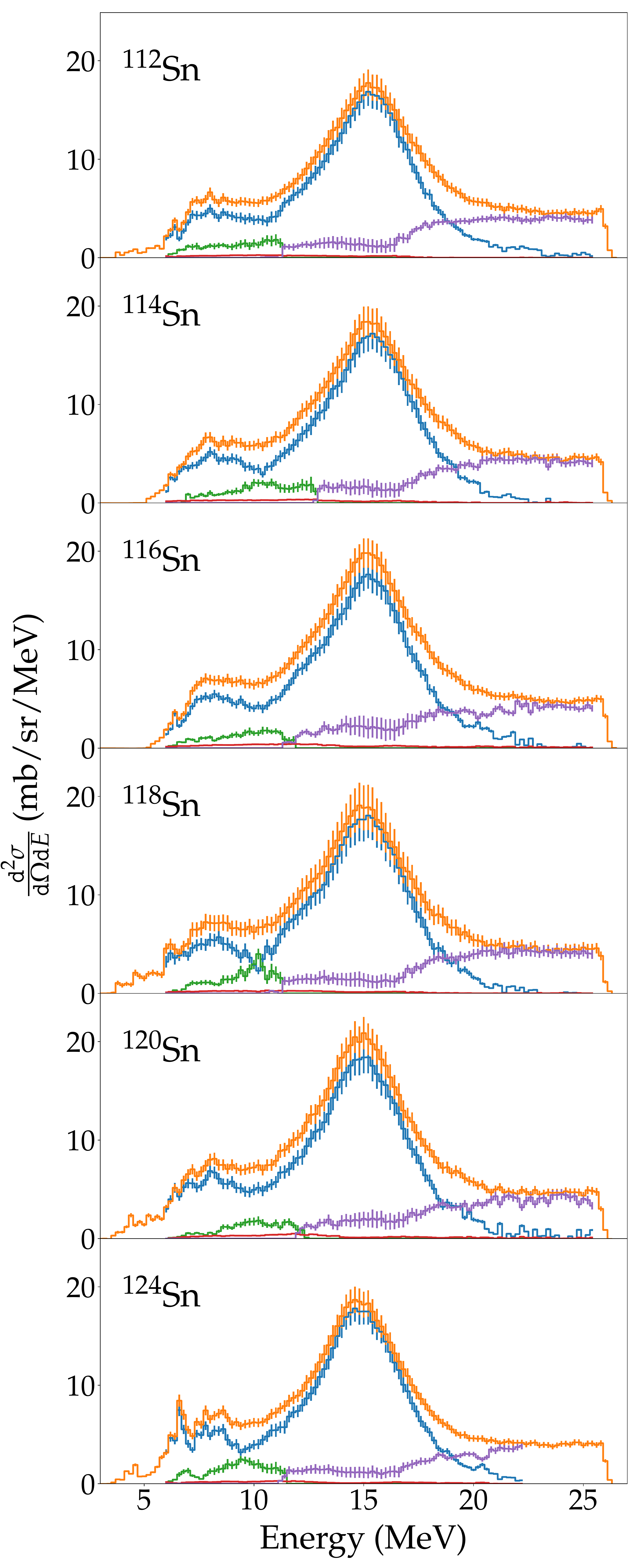}
	\caption{Results of the multipole decomposition analysis for the spectra measured at a spectrometer angle $\theta=0^\circ$.
	Orange: Experimental cross sections after subtraction of the ISGMR and ISGQR contributions. 
	Blue: $E1$ contributions. 
	Green: $M1$ contributions.
	Red: Contribtions from multipoles $\lambda > 1$.
	Purple: Continuum background.
	The error bars include statistical, systematical and MDA uncertainties.}
	\label{fig:MDAResults}
\end{figure}

The results of the MDA for all isotopes are summarized in Fig.~\ref{fig:MDAResults} presenting the full acceptance spectra measured at $0^\circ$ (cf.\ Fig.~\ref{fig:ddcs}).
The orange data show the experimental cross sections after subtraction of ISGMR and ISGQR contributions, respectively.
The error bars include statistical, systematical and MDA uncertainties added in quadrature.
The $E1$ (blue) contribution is similar in all isotopes. 
All other multipoles (red) except $M1$ (green) contribute very little.
The continuum background (purple) shows the expected increase from the neutron threshold up to the region of approximately constant cross sections above the IVGDR.
However, in the region near threshold one finds an abrupt onset at slightly different excitation energies in the different isotopes.  
Due to the similarity of the theoretical $M1$ and the continuum background angular distributions, it is difficult to distinguish these two contributions in an energy region of  \unit[$1 - 2$]{MeV} above the neutron threshold leading to a larger uncertainty of the $M1$ component not included in the error bars shown (see also Sec.~\ref{subsec5c}).

\section{Photoabsorption cross sections}
\label{sec4}

\subsection{Virtual photon method}

The conversion of Coulomb-excitation to photoabsorption cross sections is based on the virtual photon method described e.g.\ in Ref.~\cite{Bertulani1988}.
In contrast to the previous results published for $^{120}$Sn \cite{Hashimoto2015,Krumbholz2015}, which were based on the semiclassical approximation, here the virtual photon spectrum is calculated in the eikonal approximation \cite{Bertulani1993}.  
It allows for a proper treatment of relativistic and retardation effects and provides more realistic angular distributions due to taking into account absorption on a diffuse nuclear surface.
Examples of virtual photon spectra for the case of $^{120}$Sn and the differences between both approaches can be found in Sec.~3.3 of Ref.~\cite{vonNeumann-Cosel2019}.
However, the experimental data are given for an average scattering angle. 
To account for this, one needs to average the differential virtual photon number over the experimental solid angle.

\begin{figure*}
	\centering
	\includegraphics[width=0.9\textwidth]{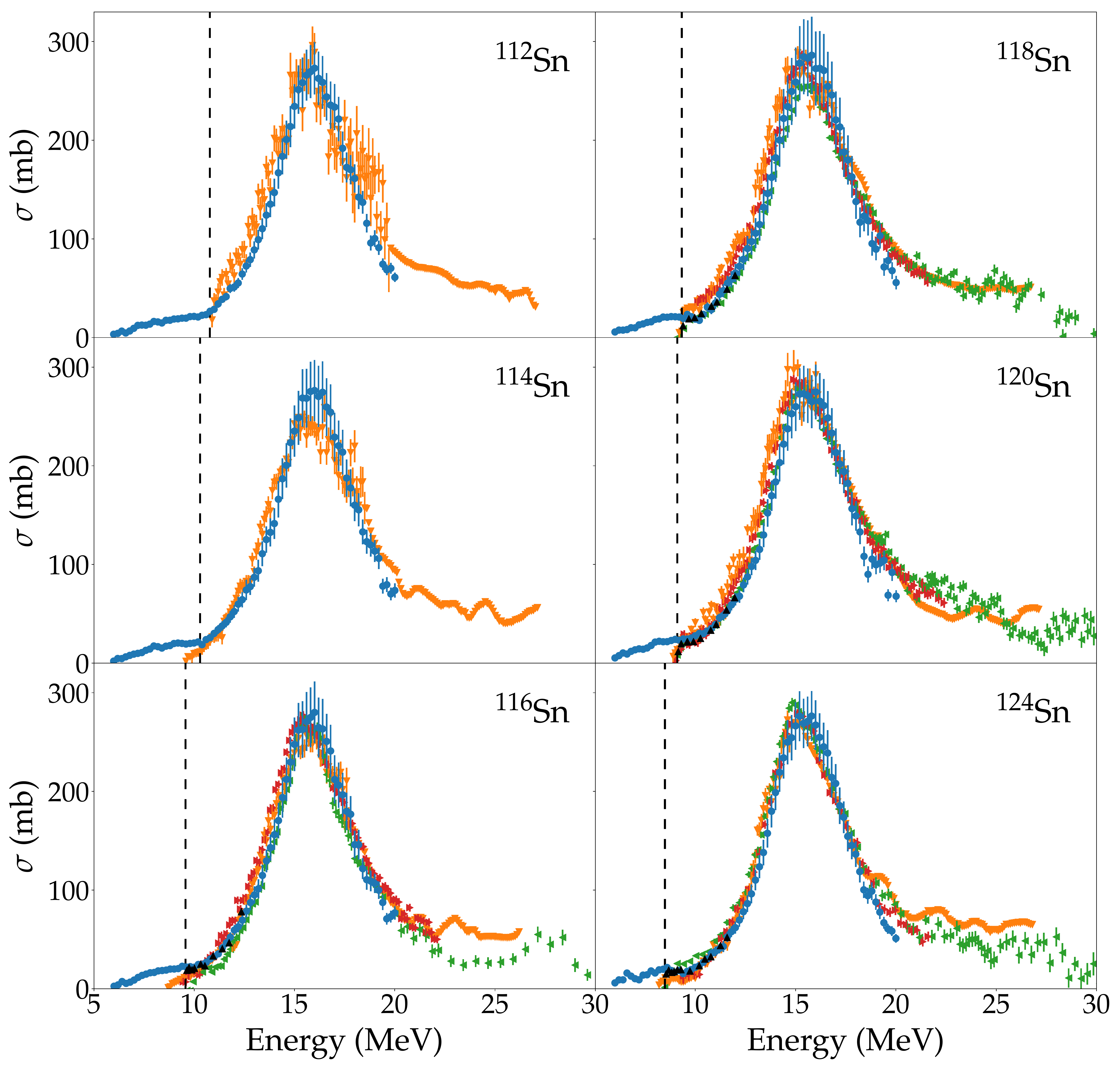}
	\caption{Photoabsorption cross sections obtained in this work (blue circles) in comparison to $(\gamma,xn)$ experiments by Fultz \textit{et al.}~\cite{Fultz1969} at Livermore (green left triangles), Leprêtre \textit{et al.}~\cite{Lepretre1974} at Saclay (red right triangles), and Sorokin \textit{et al.}~\cite{Sorokin1974, Sorokin1975} (orange downward triangles). 
	$(\gamma,n)$ data from Utsunomiya \textit{et al.}~\cite{Utsunomiya2009, Utsunomiya2011} are shown as black upward triangles. 
	The neutron thresholds are indicated by vertical dashed lines.}
	\label{fig:PhotoCS}
\end{figure*}
Another point to be considered is the maximum scattering angle at which the strong interaction between projectile and target nucleus starts to play a  role. 
This can be calculated from relativistic Rutherford scattering using \cite{Harakeh2001}
\begin{equation}
	\theta_{\rm lab}^{\rm max}=\frac{Z_1Z_2e^2}{b\mu \beta^2\gamma},
\end{equation}
where $Z_1$ is the projectile charge, $Z_2$ the charge of the target nucleus, $e$ the elementary charge, $\mu$ the reduced mass, $\beta$ the velocity in units of speed of light, $\gamma$ the Lorentz factor and $b$ the impact parameter. The impact parameter is taken as the sum of the projectile and target nucleus radii $b=r_p+r_0 A^{1/3}$, where $r_p = 0.87$ fm is the proton root mean square charge radius \cite{codata}, $r_0=\unit[1.25]{fm}$ and $A$ the mass number. 
For the investigated tin isotopes, the maximum scattering angle was determined to $\theta_{lab}^{max}=2.25^\circ-2.32^\circ$  depending on $A$.
The average differential virtual photon number is then given by
\begin{equation}
	\label{eqn:dNdOmegaMean}
	\left\langle\frac{\mathrm{d}N_{E1}}{\mathrm{d}\Omega}(E,\theta)\right\rangle = \frac{\int\frac{\mathrm{d}N_{E1}}{\mathrm{d}\Omega}(E,\theta)\mathrm{d}\Omega}{\int\mathrm{d}\Omega},
\end{equation}
where the integration is performed up to the maximum angle. 
For heavy nuclei, after integration over the relevant angular range, differences of virtual photon numbers from the semiclassical and the eikonal approach are found to be small for the present kinematics, typically less than 10\%.

\subsection{Results and comparison to previous work}
\label{subsec4b}

\begin{figure*}
	\centering
	\includegraphics[width=0.9\textwidth]{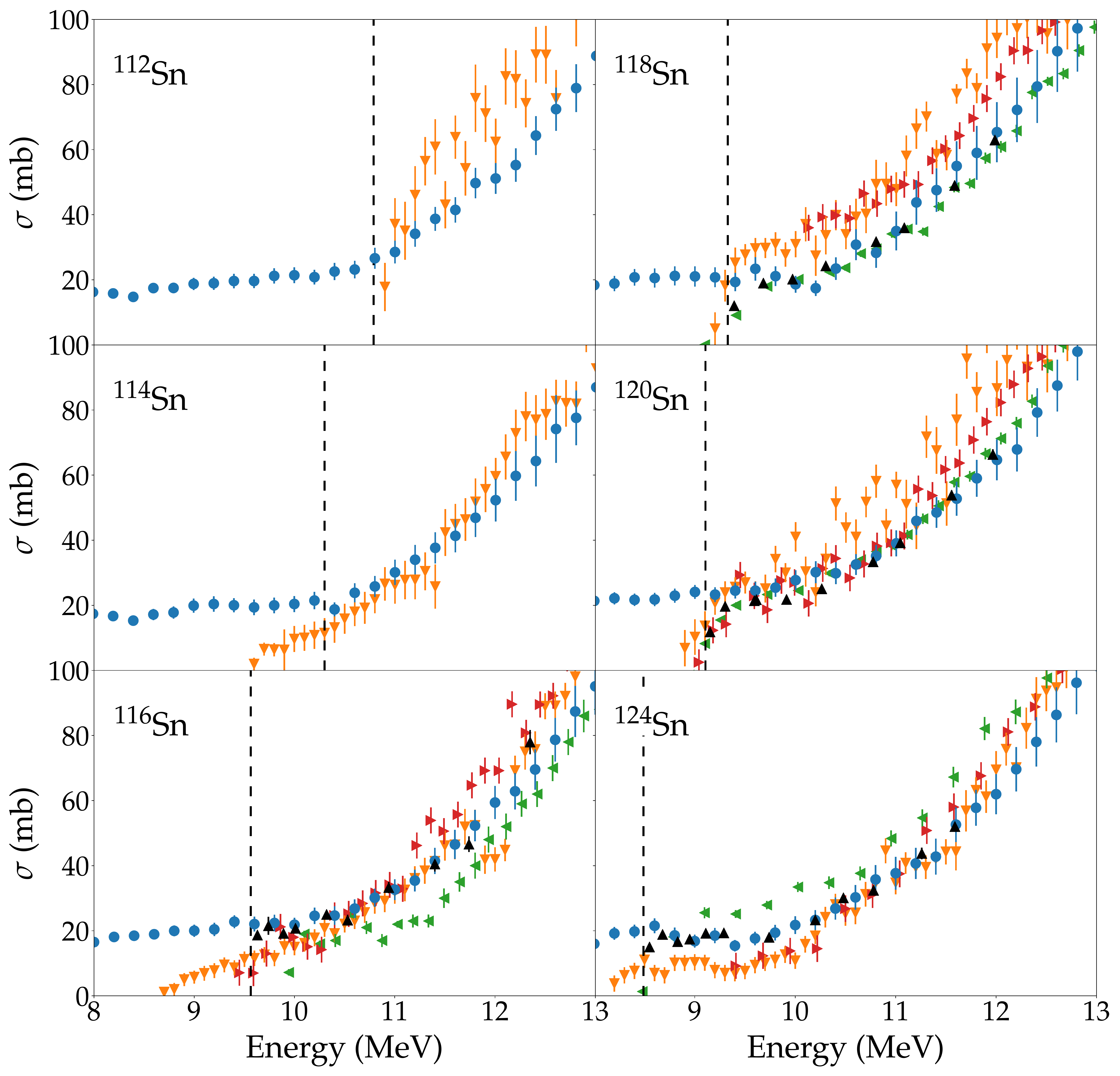}
	\caption{Same as Fig.~\ref{fig:PhotoCS}, but restricted to the energy region $8 -13$ MeV.}
	\label{fig:PhotCSlow}
\end{figure*}
The resulting photoabsorption cross sections (blue circles) are summarized in Fig.~\ref{fig:PhotoCS} in comparison to data from previous experiments.
Photoabsorption cross sections in $^{116,118,120,124}$Sn have been measured in Livermore \cite{Fultz1969} (green left triangles) and Saclay \cite{Lepretre1974} (red right triangles) with the $(\gamma,xn)$ reaction.
Additional $(\gamma,xn)$ data for all isotopes investigated here are available from Refs.~\cite{Sorokin1974,Sorokin1975} (orange downward triangles).
There are also more recent $(\gamma,n)$ data for $^{116,118,120,124}$Sn from experiments with monoenergetic photons at NEW SUBARU \cite{Utsunomiya2009,Utsunomiya2011} (black upward triangles).

In the energy region near the resonance maximum reasonable agreement is found in most cases except for the significantly lower data points of Ref.~\cite{Sorokin1975} in $^{114,116}$Sn.
Also, the Livermore results for $^{118}$Sn are below the other three experiments. 
These two data sets tend to be systematically higher than the present results on the low-energy flank of the IVGDR. 
The Saclay results, on the other hand, show a systematic relative shift with increasing $A$ from undershooting the present $^{116}$Sn results to slightly overshooting for $^{124}$Sn. 

Around the neutron threshold, however, larger deviations can be observed as illustrated in Fig.~\ref{fig:PhotCSlow}, where the energy region between 8 and 13 MeV is magnified.
The Saclay cross sections are larger than the present ones in $^{116,118,120}$Sn except close to threshold and agree for $^{124}$Sn.
The Livermore data closer to the threshold show smaller cross sections for $^{116}$Sn, cross sections similar to the present work for $^{118,120}$Sn, and larger cross sections for $^{124}$Sn.
The results of Refs.~\cite{Sorokin1974,Sorokin1975} are significantly higher for $^{112,118,120}$Sn but agree fairly well for $^{114,116,124}$Sn except the region close to threshold in $^{124}$Sn. 
On the other hand, the $(\gamma,n)$ experiments of Utsunomiya {\it et al.}~\cite{Utsunomiya2009,Utsunomiya2011} are in good agreement with the present work for all studied isotopes. 
\begin{table*}
	\centering
	\caption[Lorentz parameters for the IVGDR in $^{112,114,116,118,120,124}$Sn  from different experiments.]
			{Lorentzian fits to the IVGDR photoabsorption cross sections in $^{112,114,116,118,120,124}$Sn from different experiments.
			All data were fitted in the excitation energy range \unit[$13-18$]{MeV}.
			The results for the data of Refs.~\cite{Sorokin1974, Sorokin1975} were taken from Table 5 of Ref.~\cite{Varlamov2010}. 
			Neither uncertainties nor the fitting range are available for these numbers.}
	\begin{tabular}{lllllll}
 		\hline
 		\hline
 		\noalign{\vskip 0.5mm}
						& $^{112}$Sn & $^{114}$Sn & $^{116}$Sn 
 						& $^{118}$Sn & $^{120}$Sn & $^{124}$Sn \\
 		\hline
 		$\sigma_{\rm GDR}$ (mb) &&&&& \\
 		this work & 272(16) & 280(16) & 279(16)
 	    & 290(16) & 285(16) & 286(15) \\
 		\cite{Sorokin1974, Sorokin1975, Varlamov2010}
 		& $268$ 	 & $265$	   & $260$
 		& $272$		 & $297$	   & $270$ \\
 		\cite{Fultz1969}		
 		& $-$ 		 & $-$  	   & 266(7)
 		& 255(7) & 280(8)   & 283(8) \\
 		\cite{Lepretre1974}	
 		& $-$ 		 & $-$ 		   & 270(5)
		& 278(5)  & 284(5)  & 275(5) \\
 		\\
 		$E_{\rm GDR}$ (MeV) &&&&& \\
 		this work					
 		& 15.91(5) & 15.96(6) & 15.81(5)
 		& 15.67(8) & 15.61(5) & 15.46(5) \\
 		\cite{Sorokin1974, Sorokin1975, Varlamov2010}
 		& $15.8$		 & $15.7$ 		   & $15.6$
 		& $15.5$ 		 & $15.3$		   & $15.5$ \\
 		\cite{Fultz1969}		
 		& $-$ 		 	 & $-$  	   	   & 15.67(4)
		& 15.60(4) & 15.40(4) & 15.18(4) \\
 		\cite{Lepretre1974}	
 		& $-$ 		 	 & $-$ 		   	   & 15.57(10)
 		& 15.44(10)  & 15.38(10)  & 15.29(10) \\
 		\\
 		$\Gamma_{\rm GDR}$ (MeV) &&&&& \\
 		this work					
 		& 4.51(20) & 4.50(22) & 4.42(22)
 		& 4.47(33) & 4.48(19) & 4.33(17) \\
 		\cite{Sorokin1974, Sorokin1975, Varlamov2010}
 	    & $5.9$		    & $-$ 			 & $-$
 		& $5.8$ 		& $5.7$		     & $-$ \\
 		\cite{Fultz1969}		
 		& $-$ 		 	 & $-$  	   	 & 4.19(6)
		& 4.76(6) & 4.88(6) & 4.81(6) \\
 		\cite{Lepretre1974}	
 		& $-$ 		 	 & $-$ 		   	 & 5.21(10) 
		& 4.99(10)  & 5.25(10)  & 4.96(10)  \\
    	\hline
    	\hline
	\end{tabular}
	\label{tab:GDRParameters}
\end{table*}

At high excitation energies, the present results are shown up to 20 MeV only, since the cross section ratio between $E1$ and the conti\-nuum background becomes too unfavourable for a meaningful MDA.  
At energies in the region between 20 and 30 MeV, data are available only from Ref.~\cite{Fultz1969} and Refs.~\cite{Sorokin1974,Sorokin1975}.
The Livermore data show large variations and no isotopic trend.
The cross sections for $^{120}$Sn are about two times larger than for $^{118,124}$Sn, which in turn are larger than for $^{116}$Sn.
The data of Refs.~\cite{Sorokin1974,Sorokin1975}
are on the average more consistent with each other but show large fluctuations as a function of energy between neighboring isotopes.
These observations point towards problems in the extraction and separation of $(\gamma,2n)$ and $(\gamma,3n)$ events.

\subsection{Systematics of the IVGDR}
\label{subsec:ivgdr}

\begin{figure}
	\centering
	\includegraphics[width=\columnwidth]{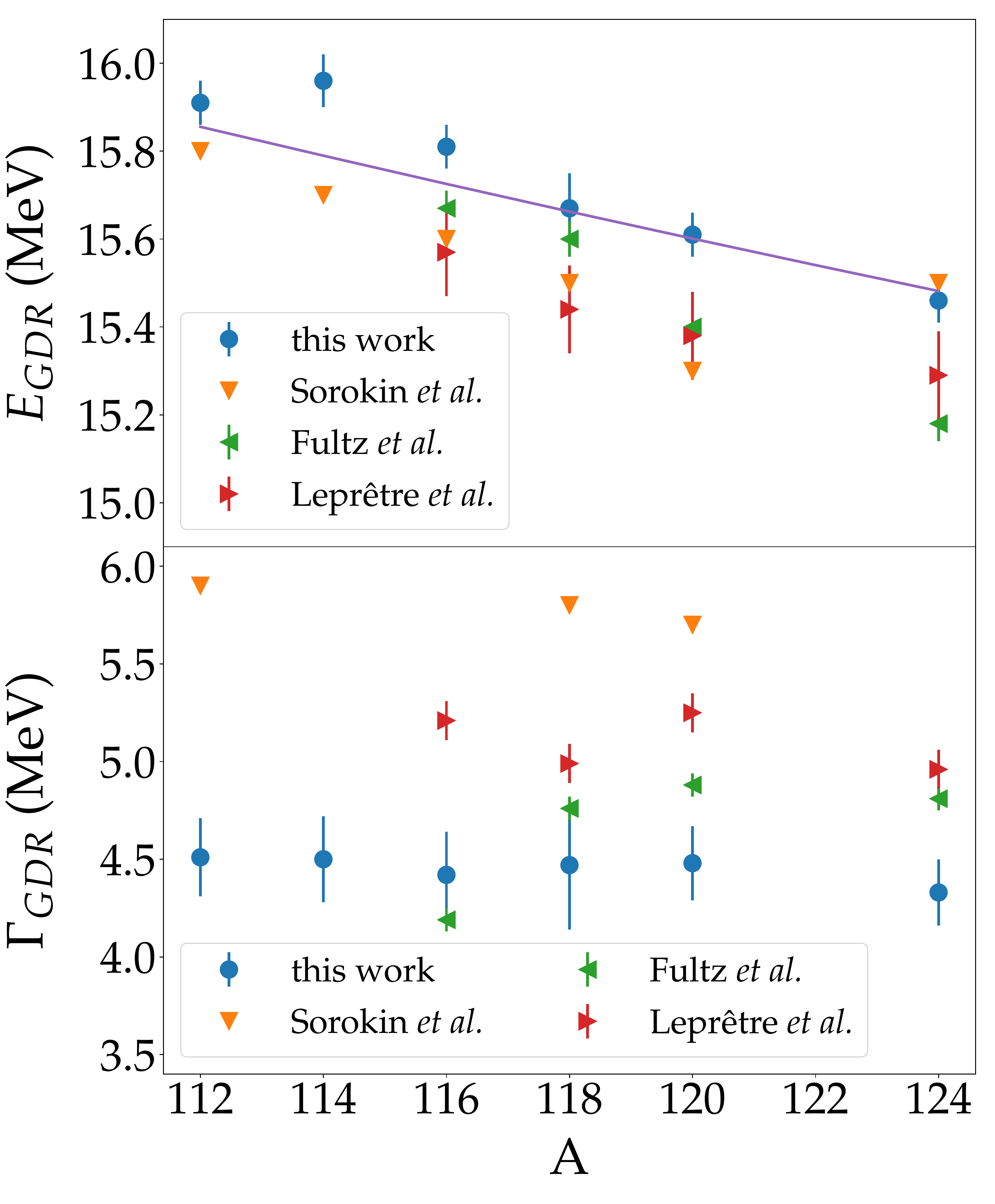}
	\caption{Centroid energies $E_{\rm GDR}$ (top) and widths $\Gamma_{\rm GDR}$ (bottom) of Lorentzian fits to the IVGDR in tin isotopes determined from the data shown in Fig.~\ref{fig:PhotoCS}.
	The purple line shows the phenomenological mass dependence of the centroid energy, Eq.~(\ref{eqn:powerlaw}).}
	\label{fig:gdrParsPhoto}
\end{figure}

Lorentzian fits to the different data sets are presented in Fig.~\ref{fig:gdrParsPhoto} and summarized in Tab.~\ref{tab:GDRParameters}.
The parameters for the present experiment were obtained using data in the energy range $13 - 18$ MeV only.
The original data of Refs.~\cite{Sorokin1974, Sorokin1975} were not accessible but fit results are given in Table 5 of Ref.~\cite{Varlamov2010}. 
Neither uncertainties nor the fitting range are available for these results.

The peak cross sections $\sigma_{GDR}$ in Tab.~\ref{tab:GDRParameters} agree very well within the uncertainties for all data sets except the aforementioned reduction of the Livermore results for $^{118}$Sn in the IVGDR peak region.
The situation is different with respect to the centroid energies and the widths as illustrated in Fig.~\ref{fig:gdrParsPhoto}. 
The expected decrease of the centroid energy $E_{\rm GDR}$ with increasing mass number $A$ is found in all experiments, though neither the absolute values nor the slope agree between the different sets of data.   
The centroid energies determined in this work are found to be generally higher than in previous work, yet they yield the best agreement comparing with the well-known phenomenological formula \cite{Berman1975}
\begin{equation}
	\label{eqn:powerlaw}
	E_{GDR} = 31.2A^{-1/3}+20.6A^{-1/6}
\end{equation}
plotted as purple line in the upper part of Fig.~\ref{fig:gdrParsPhoto}. 

The widths $\Gamma_{\rm GDR}$ differ considerably between the experiments. 
The values from the present experiment are systematically smaller. 
They are constant within the uncertainties with an average of about \unit[4.5]{MeV}. 
Likewise, the data of Fultz \textit{et al.}~\cite{Fultz1969} show a rather constant width with $E_{\rm GDR} \approx \unit[4.8]{MeV}$, except for $^{116}$Sn. 
The data of Leprêtre \textit{et al.}~\cite{Lepretre1974} exhibit a fluctuating behaviour around an average value of about \unit[5.1]{MeV}. 
The values quoted in Ref.~\cite{Varlamov2010} for the data of Refs.~\cite{Sorokin1974,Sorokin1975} are generally much larger exceeding \unit[5.5]{MeV}. \\

\subsection{IVGDR energies and nuclear matter bulk parameters}

The tool of choice for the theoretical modeling of nuclear giant resonances is since long the Random-Phase Approximation (RPA). 
It is based on a mean-field description in terms of one-particle-one-hole states recoupled by a residual two-body interaction \cite{Rin80aB}.
Early realizations of RPA were mostly based on empirical shell-model potentials and separately added interactions
\cite{Goeke1982}.  
Meanwhile, steady progress in nuclear energy density functional theory \cite{Bender2003} and in numerical capabilities has made fully self-consistent RPA calculations a widely used standard tool.  
However, most EDFs are tuned to ground state properties which leaves its isovector properties to some extent undetermined. 
The RPA predictions for the IVGDR are thus widely varying.
A proper tuning of the IVGDR within nuclear EDF theory is a field of active research, see e.g., Refs.~\cite{Paa07aR,Kluepfel2009,Roca-Maza2018}, and any precise new data are highly welcome. 

Thus we will now compare the present measurements
with a variety of RPA predictions. 
To do that in systematic manner, we chose a family of EDF parameterizations which vary certain nuclear-matter properties (NMP) in systematic manner, i.e.,  they all describe the same pool of ground-state properties equally well, but differ in one of the NMP varied within acceptable bounds while leaving the quality of ground state properties intact. 
There exist several such sets of families from Skyrme EDF as well as from relativistic models \cite{Agrawal2005,Kluepfel2009,Nazarewicz2014,Yuksel2019}.  
We confine the present study to the set from Ref.~\cite{Kluepfel2009} as it covers the broadest set of NMP (see below). 
For $^{208}$Pb a one-to-one relation between each major giant resonance and one of the NMP was found \cite{Kluepfel2009} and corroborated by statistical correlation analysis \cite{Erler2015}. 
On the other hand, the quality of the description of IVGDR can change dramatically with nuclear size \cite{Erl10a}. 
The tin isotopes are about half the size of lead and it is interesting to see how the RPA description performs in this case.

RPA is capable to describe the gross properties of giant resonances well, but it fails if one looks at the detailed profile of the strength distributions. 
RPA spectra for the IVGDR are too much structured while experimental data show usually one broad GDR peak, see e.g. Fig.~\ref{fig:PhotoCS}. 
It requires two-body correlations beyond RPA to describe the spreading width \cite{Ber83aR}.  
Practical realizations in terms of the phonon-coupling model are, indeed, able to produce realistically smooth excitation spectra \cite{Col01a,Tse16a}.
These are, however, very expensive to use and contain too many ingredients for a simple comparison with data. 
Before going into details, one has first to check the gross properties and this can be done very well at the simpler level of RPA when comparing averaged properties. 
One such quantity is the dipole polarizability discussed in Sec.~\ref{subsec5b} and Ref.~\cite{Bassauer2020} which can be obtained from the $(-2)$ moment of the photoabsorption cross section distribution. 
The other prominent feature is the IVGDR peak position scrutinized here. 
A quick glance at Fig.~\ref{fig:PhotoCS} shows that 
it is unsafe to read off the peak position directly from the strength distribution. 
A more robust value is obtained from the average energy in a given interval $[E_1,E_2]$
\begin{equation}
  \overline{E}
  =
  \frac{\int_{E_1}^{E_2}dE\sigma_\mathrm{D}(E)E}
       {\int_{E_1}^{E_2}dE\sigma_\mathrm{D}(E)} \,.
\label{eq:peaks}
\end{equation}
The Lorentzian fits to the data in Sec.~\ref{subsec4b} have been performed for an energy region \unit[$13-18$]{MeV} and the same interval is chosen for the  theoretical results. 
Before doing that, the RPA spectra are folded to resemble approximately the smoothness of the data.  
To explore the impact of smoothing, we have used Lorentzian as well as Gaussian folding with widths from 1 to \unit[2]{MeV}. 
The results are found to be only weakly dependent on the actual folding recipe.  
We take the variations of the resulting peak energies as uncertainties of the analysis, shown as error bars in the following.

\begin{figure*}
	\centering
	\includegraphics[width=0.8\textwidth]{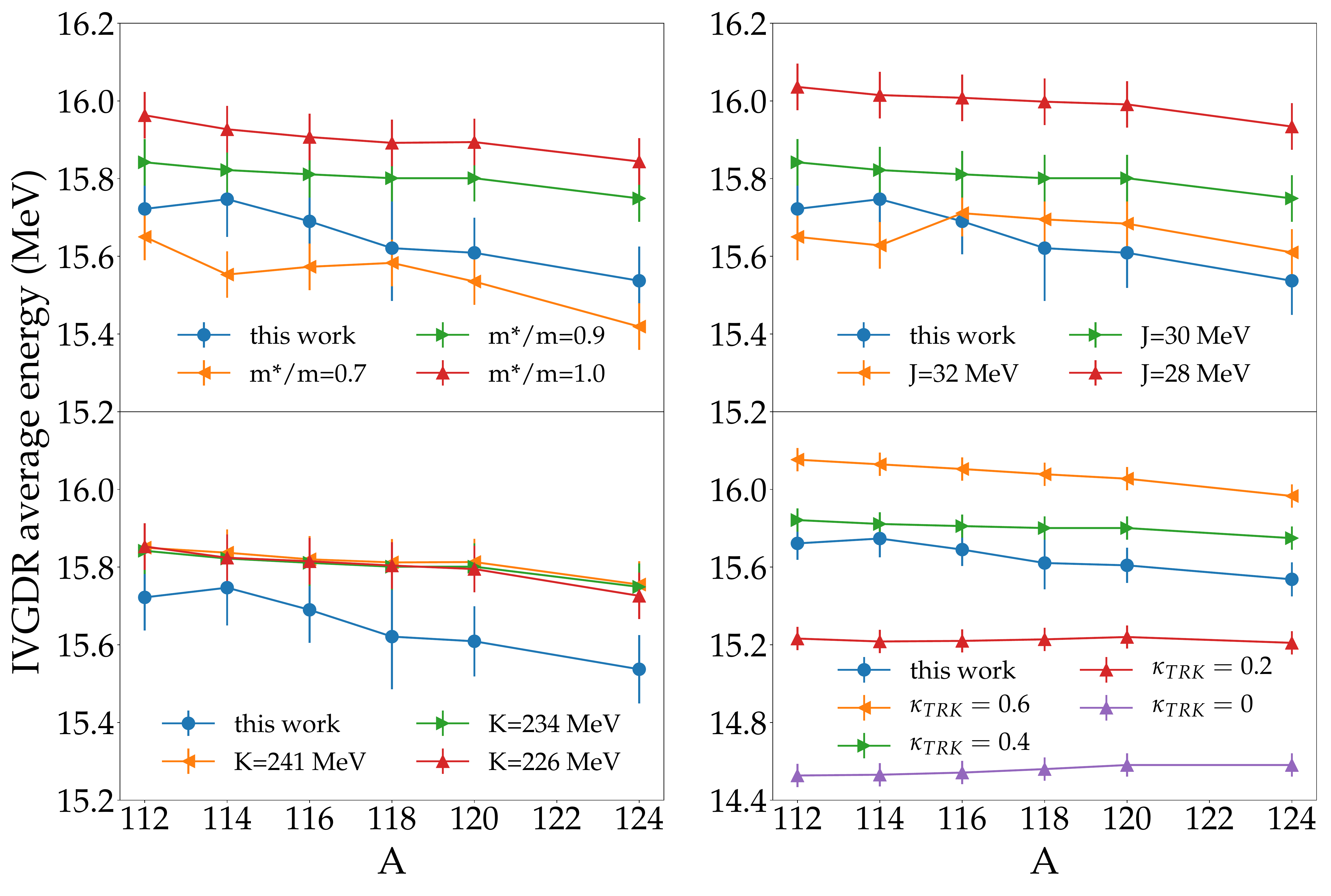}
	\caption{\label{fig:Sn-GDR-aver1318} Average IVGDR peak positions, Eq.~(\ref{eq:peaks}), along the chain of tin isotopes. 
	Compared are experimental data (blue) with results from various Skyrme parameterizations as indicated (for details see text).  
	Each panel collects variation of one NMP. 
	The results of the original SV-bas interaction \cite{Kluepfel2009} are always shown in green.}
	\label{fig:IVGDR-aver1318}
\end{figure*}
\begin{table}
\caption{\label{tab:nucmat} 
NMP for the family of Skyrme parameterizations from \cite{Kluepfel2009}, where SV-bas is the base point of the systematic variation. 
Variations of the incompressibility $K$ are given in SV-K*, of the effective mass $m^*/m$ in SV-mas*, of the symmetry energy $J$ in SV-sym*, and of the TRK sum rule enhancement $\kappa_\mathrm{TRK}$ in SV-kap*.}
\begin{tabular}{lcccc}
\hline
\hline
force & $K$ & $m^*/m$ & $a_\mathrm{sym}$ & $\kappa_\mathrm{TRK}$  \\
& (MeV) & & (MeV) &   \\
\hline
SV-bas   &  234 &  0.9 & 30   & 0.4 \\
\hline
SV-K218 &  218 &  0.9 & 30   & 0.4 \\
SV-K226 &  226 &  0.9 & 30   & 0.4 \\
SV-K241 &  241 &  0.9 & 30   & 0.4 \\
\hline
SV-mas10 &  234 &  1.0 & 33   & 0.4  \\
SV-mas08 &  234 &  0.8 & 26   & 0.4 \\
SV-mas07 &  234 &  0.7 & 20   & 0.4 \\
\hline
SV-sym28 &  234 &  0.9 & 28   & 0.4 \\
SV-sym32 &  234 &  0.9 & 32   & 0.4 \\
SV-sym34 &  234 &  0.9 & 34   & 0.4 \\
\hline
SV-kap00 &  234 &  0.9 & 30   & 0.0 \\
SV-kap20 &  234 &  0.9 & 30   & 0.2 \\
SV-kap60 &  234 &  0.9 & 30   & 0.6 \\
\hline
\hline
\end{tabular}
\end{table}

As said above, we compare data with RPA results from a family of Skyrme functionals dervied from SV-bas \cite{Kluepfel2009}, which varies systematically the four crucial NMP: incompressibility $K$, symmetry energy $J$, isoscalar effective mass $m^*/m$, and isovector effective mass in terms of the Thomas-Reiche-Kuhn sum rule enhancement $\kappa_\mathrm{TRK}$.
The NMP for all parameterizations are given in Tab.~\ref{tab:nucmat}.
SV-bas was developed as a Skyrme functional which fits the
three major giant resonances and the dipole polarizability in
$^{208}$Pb together with an excellent description of ground state properties.  
For $^{208}$Pb, it was found that each NMP has a one-to-one
correlation with one giant resonance peak energy \cite{Kluepfel2009,Erler2015}, viz.\ the ISGMR with $K$,
the ISGQR with $m^*/m$, the IVGDR with $\kappa_\mathrm{TRK}$, and the dipole polarizability $\alpha_D$ with $J$.
This means, e.g., for the IVGDR peak that variation of $K$,
$m^*/m$, and $\alpha_D$ has negligible effect while
$\kappa_\mathrm{TRK}$ has direct impact on the result. 
The question is how the IVGDR in the tin isotopes behaves in that respect.

Figure \ref{fig:Sn-GDR-aver1318} compares the average IVGDR peak energies from the various Skyrme parameterizations with those from the strength distributions of the present experiment.  
At first glance, the behavior is similar to what we have seen in $^{208}$Pb: SV-bas is still fairly well fitting, variation of $\kappa_\mathrm{TRK}$ has a strong impact, and the other variations change less.  
Closer inspection, however, reveals remarked deviations from the simple behavior in $^{208}$Pb. 
First, variation of $J$ and $m^*/m$ is not totally inert (as $K$ still is), but has some impact, indicating that the near perfect one-to-one correlation between a giant resonance and ``its'' NMP is weakened in the tin isotopes. 
Second, SV-bas predicts \unit[$100 - 200$]{keV} higher centroid energies than seen experimentally, while the description is almost perfect in $^{208}$Pb.  
This indicates that the mass dependence of isovector properties is not yet fully modeled by present-day EDFs, much in line with our findings for the isovector polarizability \cite{Bassauer2020} and earlier
studies of the $A$-dependence of the IVGDR \cite{Erl10a}. 

The isotopic trend of the data was found to agree with the known phenomenological form, Eq.~(\ref{eqn:powerlaw}). 
Most of the theoretical results comply with this trend.
Only the variation of $\kappa_\mathrm{TRK}$ shows slight changes, but these fine details go beyond the resolution of the present analysis and data.  
The results along an isotopic chain thus confirm the known trend of IVGDR with mass number $A$. 
On the other hand, the large step from $A=208$ to $A\approx 120$ revealed deviations. 
It is not yet clear whether this is due to the larger change in $A$ or due to a change in charge number $Z$. 
The present data are thus one important entry for a future
systematic study.

\section{Electric and magnetic dipole strength}
\label{sec5}

In this section the systematics of the $E1$ and $M1$ strength distributions in the studied tin isotopes is discussed.
The $B(E1)$ strength distributions are derived from the photoabsorption cross sections. 
In Ref.~\cite{Birkhan2016} a method to extract the spin-$M1$ strength from the $M1$ cross sections in forward-angle $(p,p^\prime)$ experiments has been introduced and successfully tested.
Under the assumption that isoscalar and orbital contributions can be neglected one can convert the results to the equivalent electromagnetic $B(M1)$ strength.
This works particularly well for magic nuclei \cite{Birkhan2016,Mathy2017}.
Since the magnitude of orbital contributions is related to the ground-state deformation \cite{Heyde2010}, this should also be a good approximation for the semimagic tin isotopes.

\subsection{$E1$ strength below the neutron threshold}

\begin{figure}
	\centering
	\includegraphics[width=\columnwidth]{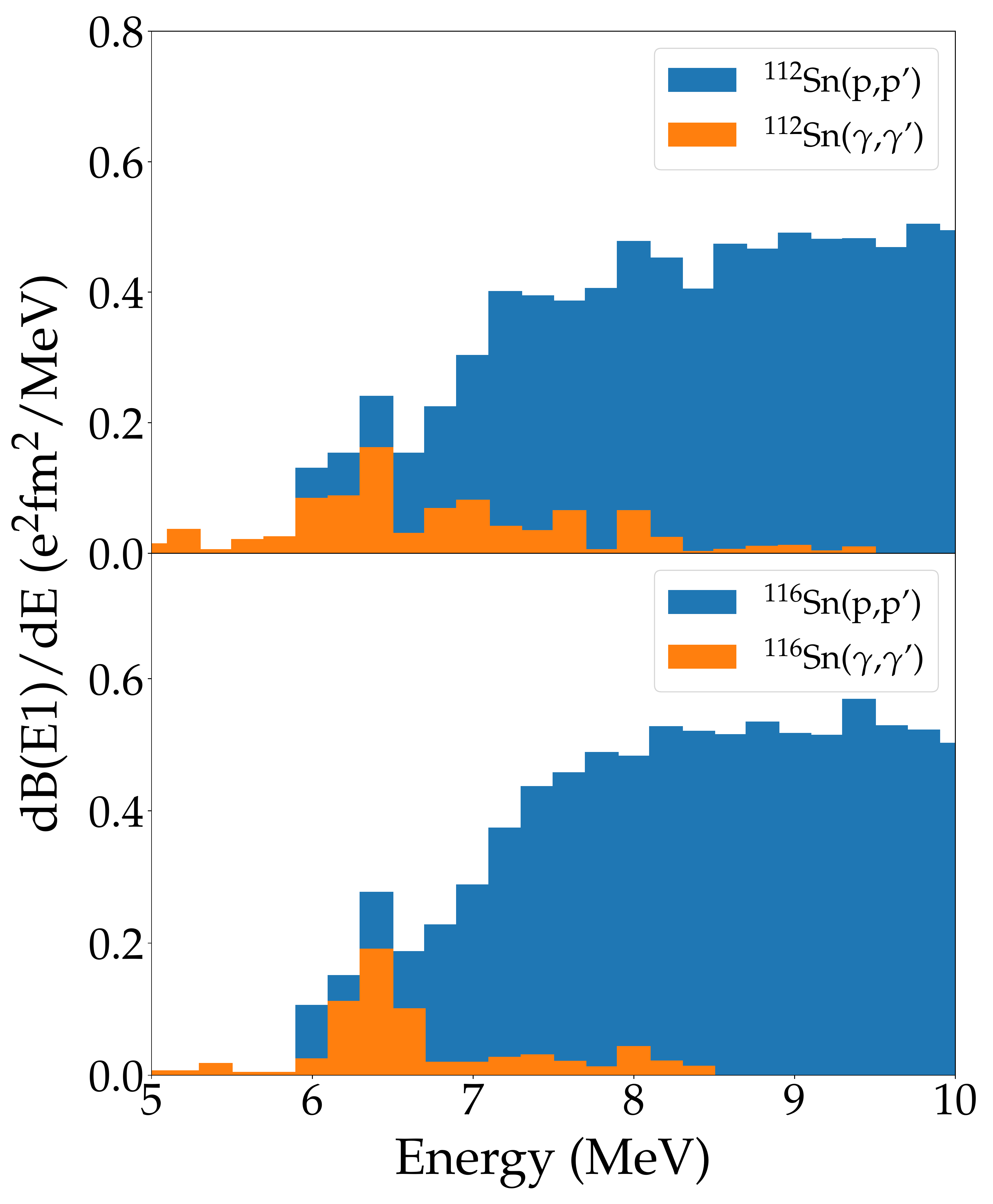}
	\caption{$B(E1)$ strength distributions for $^{112}$Sn and $^{116}$Sn below the neutron threshold in \unit[200]{keV} bins from the present work (blue) in  comparison with results from NRF experiments~\cite{Oezel-Tashenov2014, Govaert1998} (orange).
	}
	\label{fig:112116SnBE1}
\end{figure}

Below the neutron threshold, comparison can be made with data from nuclear resonance fluorescence (NRF) experiments. 
Strength distributions from NRF experiments are available for $^{112,116,120,124}$Sn \cite{Oezel-Tashenov2014, Govaert1998}. 
For $^{120}$Sn, a comparison between the $B(E1)$ strengths deduced from proton scattering and from NRF data was presented already in Ref.~\cite{Krumbholz2015}. 
It was found that in proton scattering considerably more $E1$ strength is observed, in particular at energies close to the neutron threshold. 
In Fig.~\ref{fig:112116SnBE1}, the same comparison is shown for $^{112}$Sn and $^{116}$Sn. 
As in $^{120}$Sn, an approximate agreement is seen in the region up to \unit[6.5]{MeV}, in particular if inelastic branchings (estimated with statistical model calculations) are included for the NRF data~\cite{Krumbholz2015}. 
Above \unit[6.5]{MeV}, substantially more strength is found for both isotopes measured in proton scattering. 

There are two potential explanations for these findings.
Due to the high level density in the tin isotopes much of the strength cannot be resolved in NRF experiments close to the neutron thresholds~\cite{Schwengner2007, Savran2008}, which leads to lower $B(E1)$ values. 
Furthermore, excitation strengths are usually determined under the assumption that decays to excited states are negligible. 
This assumption however is not always justified~\cite{Isaak2013, Loeher2016} and can lead to a severe underestimation of the $B(E1)$ strength. 

\begin{figure}
	\centering
	\includegraphics[width=\columnwidth]{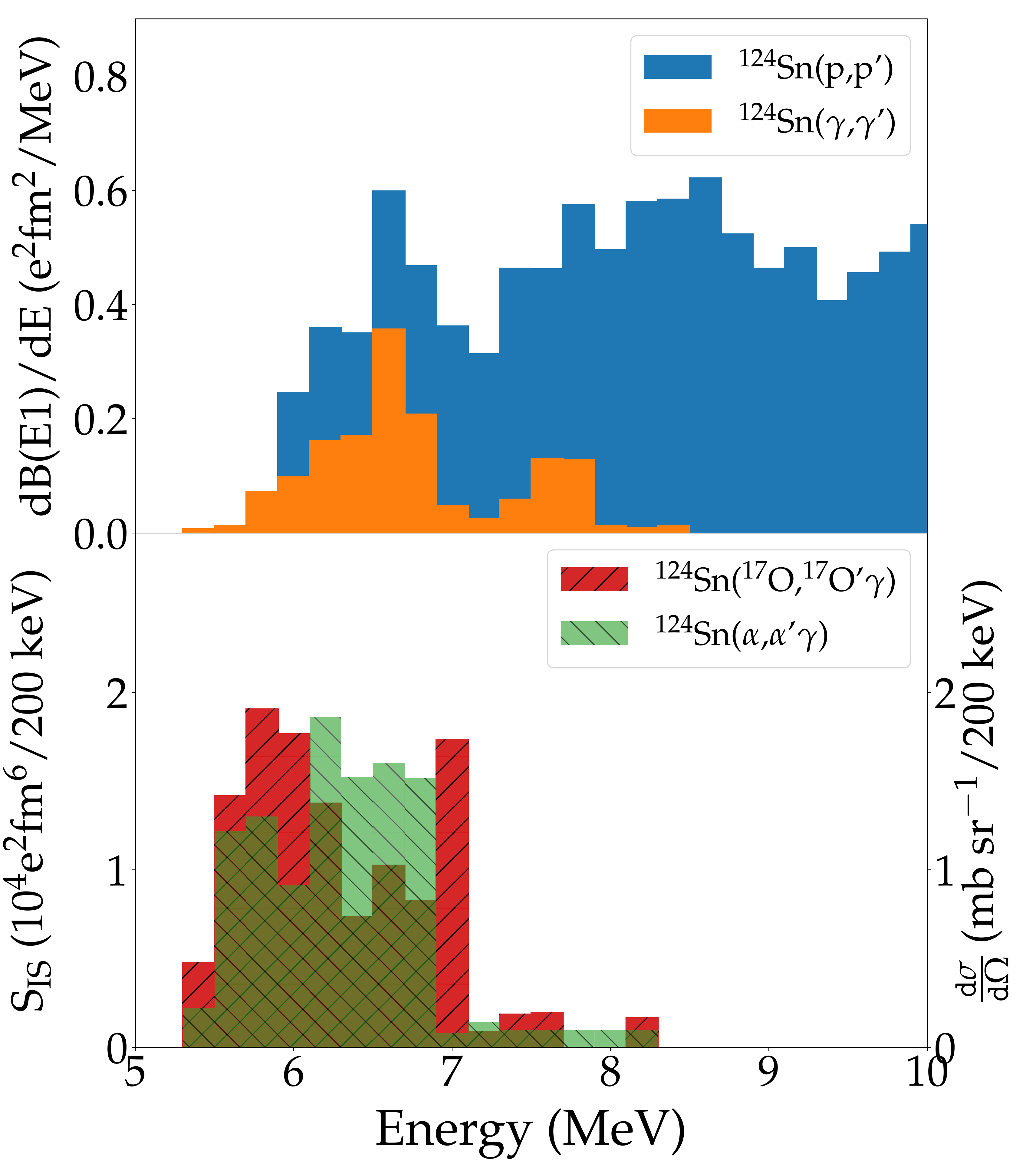}
	\caption{Electric dipole strength distributions for $^{124}$Sn in \unit[200]{keV} bins from different experiments.
	Top: $B(E1)$ strength distributions for $^{124}$Sn from the present work (blue) in comparison with NRF results~\cite{Govaert1998} (orange).
	Bottom: Isoscalar $E1$ strength distributions deduced from  ($^{17}$O,$^{17}$O$^\prime\gamma$) experiment~\cite{Pellegri2014} (red) and differential cross sections from an ($\alpha,\alpha^\prime\gamma$) experiment~\cite{Endres2012} (green). 
	}
	\label{fig:124SnBE1}
\end{figure}

Figure~\ref{fig:124SnBE1} shows results from four different experiments studying the electric dipole response in $^{124}$Sn. 
While the present $(p,p^\prime)$ and ($\gamma,\gamma$') experiments induce predominantly isovector transitions, the
($^{17}$O,$^{17}$O$^\prime\gamma$) \cite{Pellegri2014} and ($\alpha,\alpha^\prime\gamma$) \cite{Endres2012} studies probe the isoscalar response. 
As in the cases of $^{112,116,120}$Sn, a strong increase of the $B(E1)$  strength is found towards excitation energies \unit[$> 7$]{MeV} in the proton scattering data of $^{124}$Sn in contrast to the NRF data. The structure around \unit[6.5]{MeV} observed in lighter tin isotopes is even more prominent in $^{124}$Sn and clearly seen in both experiments. 
A completely different picture results from the ($^{17}$O,$^{17}$O'$\gamma$) and ($\alpha$,$\alpha$'$\gamma$) experiments A comparable isoscalar $E1$ response is found in the energy region \unit[$5.5-7$]{MeV}, although the distribution differs in detail. 
No isoscalar $E1$ matrix elements were extracted from the $\alpha$ scattering data, thus no quantitative comparison is possible.

Above \unit[7]{MeV}, hardly any isoscalar $E1$ strength is found.
From the macroscopic picture describing the PDR as a neutron-skin oscillation against an isospin saturated core, one expects mixed isoscalar and isovector excitation as seen in all four experiments below \unit[7]{MeV}. 
Absence of isoscalar strength at higher energies indicates that the strength observed in the present experiment arises from the low-energy tail of the IVGDR.
Such transitions are expected to involve complex wave functions with potentially small branching ratios to the ground state, which might explain the absence of these transitions in the NRF experiments.

\subsection{Dipole polarizability}
\label{subsec5b}

The electric dipole polarizability $\alpha_{\rm D}$ of a nucleus is related to the photoabsorption cross sections, respectively the $B(E1)$ strength distributions by inverse moments of the $E1$ sum rule \cite{Bohigas1981}
\begin{equation}
\label{eq:pol}
  \alpha_\mathrm{D}
  =
  \frac{\hbar c}{2\pi^{2} } 
  \int \frac{\sigma_\mathrm{abs}}{E_{\rm x}^{2}}{\rm d}E_{\rm x} 
  = 
   \frac{8 \pi}{9} \int\frac{\rm{B(E1)}}{E_{\rm x}}{\rm d}E_{\rm x}.
\end{equation}
The present data provide photoabsorption cross sections for the determination of $\alpha_{\rm D}$ in the energy region \unit[$6 -20$]{MeV} as discussed in Sec.~\ref{sec4}.
Below \unit[6]{MeV}, $B(E1)$ strength distributions of $^{112,116,120,124}$Sn have been measured in $(\gamma,\gamma^\prime)$ experiments  \cite{Oezel-Tashenov2014,Govaert1998}.
These contributions are small (\unit[$<0.5$]{\%} of the total dipole polarizability) and were neglected for consistency with the other isotopes, where no such data are available. 
 
In Ref.~\cite{Roca-Maza2015} it was pointed out that the quasideuteron mechanism \cite{Schelhaas1988} dominates the photoabsorption for high excitation energies (above \unit[30]{MeV} in the present case).  
Such a nonresonant process is not included in the EDF calculations and should thus be excluded from the integration of Eq.~(\ref{eq:pol}) for a comparison with theoretical results.
Rather we employ a theory-assisted estimate of strength in the region above \unit[20]{MeV} based on Quasiparticle Phonon Model (QPM) calculations known to account well for properties of the IVGDR in heavy nuclei \cite{vonNeumann-Cosel2019,Ryezayeva2002,Poltoratska2012,Tamii2011,Poltoratska2014}.  
The QPM cross sections used to calculate the dipole polarizability in the energy region above \unit[20]{MeV} were convoluted with Lorentzians whose widths were tuned to reproduce the present IVGDR data.
In order to estimate the model dependence of this procedure, the analysis was repeated with other EDF parameterizations and the predicted contributions were found to be all similar.
The upper limit of the integration was chosen as \unit[50]{MeV}, which roughly corresponds to the single-particle model space of the theoretical results.
For further details, see. Ref.~\cite{Bassauer2020}.

\begin{figure}
	\centering
	\includegraphics[width=0.97\columnwidth]{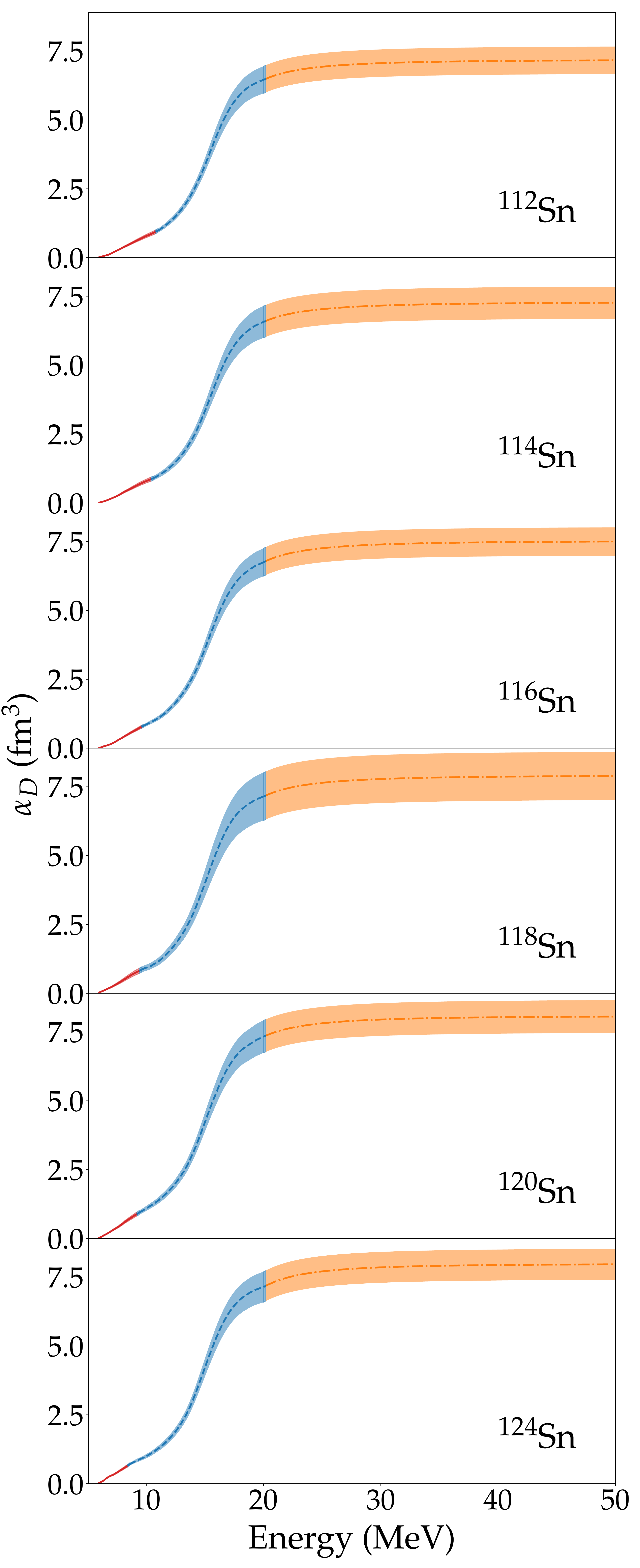}
	\caption{Running sums of the dipole polarizability deduced from the present (p,p$^\prime$) data.
	Red: Contribution from 6 MeV to $S_{\rm n}$. 
	Blue: Contribution from $S_{\rm n}$ to \unit[20]{MeV}.
	Orange: Contribution above 20 MeV from QPM calculations, see text for details.
	}
	\label{fig:DPRunSum}
\end{figure}
Figure \ref{fig:DPRunSum} displays the evolution of $\alpha_{\rm D}$ as a function of excitation energy (the running sum) for the investigated tin isotopes.  
The error bands consider statistical and systematical uncertainties, the latter including contributions from experiment and from the MDA as discussed above.
Note that the relative errors are similar at low and high excitation energies.
The figure illustrates that the total polarizabilities are dominated by the contribution of the IVGDR (blue), but the low-energy (red) and high-energy (orange) parts are non-negligible.
The corresponding total and partial values are summarized in Table \ref{tab:pol}.  
The variation of the low-energy contribution up to the neutron threshold ($S_{\rm n}$) -- i.e.\ the part missed in ($\gamma$,xn) experiments -- is driven by two counteracting factors, viz.\ the decrease of the IVGDR centroid energy and of $S_n$ with increasing $A$.
The variation of $S_n$ between $^{112}$Sn and $^{124}$Sn is more than \unit[2]{MeV} (cf. Tab.~\ref{tab:pol}).
Since the variation of the IVGDR centroid energy is only about \unit[0.5]{MeV} and the IVGDR widths are approximately constant (cf.\ Tab.~\ref{tab:GDRParameters}), the largest contribution of \unit[13]{\%} is found in $^{112}$Sn dropping to \unit[8]{\%} in $^{124}$Sn.
The high-energy contribution from the QPM calculations amounts to \unit[$9-10$]{\%} in all isotopes.
\begin{table}
	\centering
	\caption{Total dipole polarizability $\alpha_{\rm D}$ of $^{112,114,116,118,120,124}$Sn determined as described in the text.
	Partial values are given for the contributions from \unit[6]{MeV} to the neutron threshold energies $S_{\rm n}$ given in the first column, from $S_{\rm n}$ to \unit[20]{MeV}, and \unit[$> 20$]{MeV}.
	}
	\begin{tabular}{cccccc}
 		\hline
 		\hline
 		\noalign{\vskip 0.5mm}
 		& $S_n$ (MeV) & \multicolumn{4}{c}{$\alpha_{\rm D}$ (fm$^3$)} \\
 		&  & $6-S_{\rm n}$ & $S_{\rm n}-20$ & $>20$ & Total  \\
 		\hline
 		\noalign{\vskip 0.5mm}
 		$^{112}$Sn & 10.79 & $0.94(7)$ & $5.51(42)$ & $0.73(7)$ & $7.19(50)$ \\
 		$^{114}$Sn & 10.30 & $0.83(7)$ & $5.74(51)$ & $0.72(7)$ & $7.29(58)$ \\
 		$^{116}$Sn & 9.56  & $0.77(6)$ & $5.98(45)$ & $0.77(8)$ & $7.52(51)$ \\
 		$^{118}$Sn & 9.32  & $0.78(9)$ & $6.36(78)$ & $0.77(8)$ & $7.91(87)$ \\
 		$^{120}$Sn & 9.10  & $0.84(7)$ & $6.49(52)$ & $0.75(8)$ & $8.08(60)$ \\
 		$^{124}$Sn & 8.49  & $0.65(5)$ & $6.49(51)$ & $0.85(8)$ & $7.99(56)$ \\
    	\hline
    	\hline
	\end{tabular}
	\label{tab:pol}
\end{table}

\begin{figure}
	\centering
	\includegraphics[width=\columnwidth]{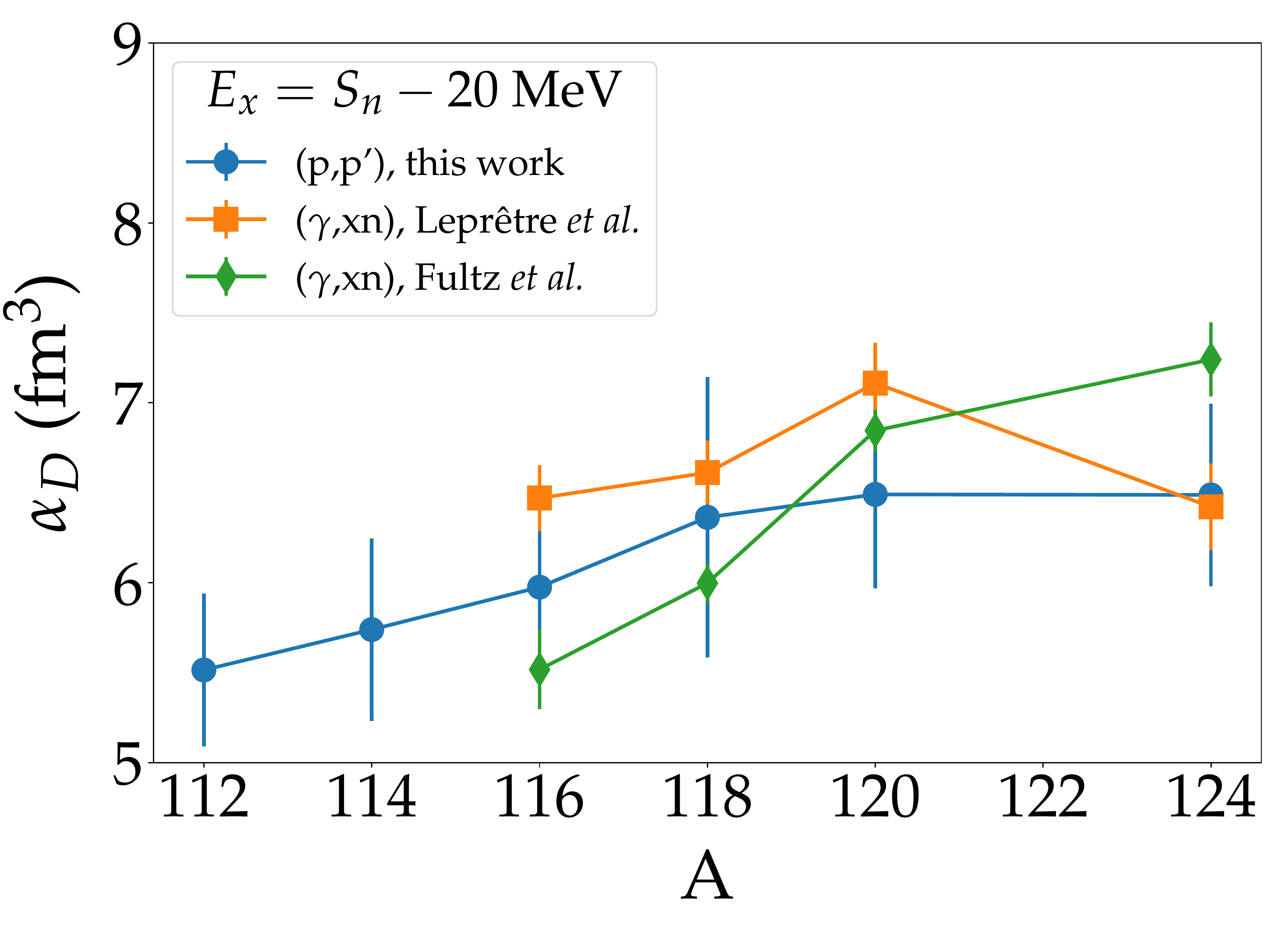}
	\caption{
	Contribution to the dipole polarizability in the energy region from $S_n$ to \unit[20]{MeV} deduced from the present data (blue circles), Ref.~\cite{Fultz1969} (green diamonds), and Ref.~\cite{Lepretre1974} (orange squares).   
	\label{fig:alphad-gdr}
	}
\end{figure}

\begin{figure}
	\centering
	\includegraphics[width=\columnwidth]{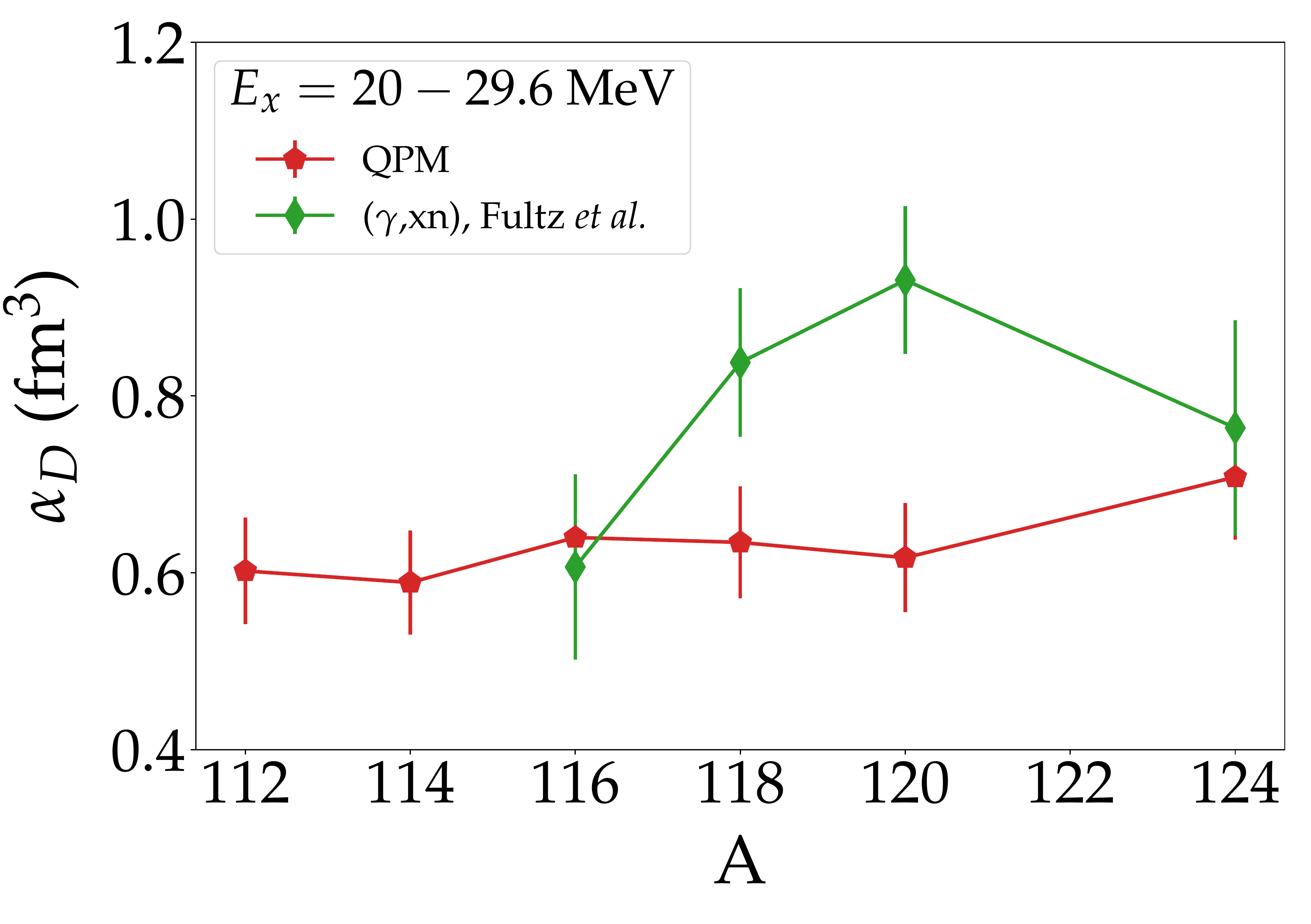}
	\caption{
	Dipole polarizability of stable even-mass tin isotopes in the energy region from 20 to \unit[29.6]{MeV} deduced from the data of Ref.~\cite{Fultz1969} (green diamonds) compared with the theory-based estimate used for the present results (red pentagons).    
	\label{fig:alphad-highenergy}
	}
\end{figure}

Above neutron thresholds, results are also available from $(\gamma,xn)$ experiments \cite{Fultz1969,Lepretre1974,Sorokin1974,Sorokin1975}, which in principle allow to reduce the error bars by averaging over energy regions covered by more than one experiment or not covered by the present data.
However, we refrain from using them, since they show large variations between different isotopes and systematically different isotopic dependence as discussed in Sec.~\ref{subsec4b} and illustrated in Figs.~\ref{fig:alphad-gdr} and \ref{fig:alphad-highenergy}. 
Figure \ref{fig:alphad-gdr} compares the polarizabilites deduced by the different experiments in the energy region from the neutron threshold to \unit[20]{MeV}.
It is obvious that if one would include the data of Ref.~\cite{Fultz1969}, the isotopic dependence of $\alpha_{\rm D}$ would be changed significantly.
Concerning the isotopic dependence extracted from the data of Ref.~\cite{Lepretre1974} one should note that for $^{118,124}$Sn results are available only from about \unit[1]{MeV} above $S_n$.  

Data for $^{116,118,120,124}$Sn in the excitation energy region \unit[$20 -30$]{MeV} not covered in the present experiments are available from Ref.~\cite{Fultz1969}. 
However, these results again show large variations between different isotopes and no systematic isotopic dependence as illustrated in Fig.~\ref{fig:alphad-highenergy}.
The problems are aggravated looking at the energy dependence of the photoabsorption cross sections. 
Between 20 and \unit[25]{MeV}, they are about two times smaller for $^{116}$Sn than those for $^{118,124}$Sn, which in turn are significantly smaller than those for $^{120}$Sn.
On the other hand, between 26 and \unit[28]{MeV}, the cross sections for $^{116}$Sn are about two times larger than those for $^{120}$Sn

We note that a larger $\alpha_{\rm D}$ value was published for $^{120}$Sn based on the same type of experiment \cite{Hashimoto2015}, which after correction for the quasideuteron part amounted to  \unit[$\alpha_\mathrm{D} = 8.59(37)$]{fm$^3$}.
However, as pointed out in Sec.~\ref{sec2} the difference to the present result is not due to the (p,p$^\prime$) data (cross sections from the previous and present experiments agree within error bars). 
Rather they result from averaging with the ($\gamma$,xn) data of Refs.~\cite{Fultz1969,Lepretre1974}, whose contributions to $\alpha_{\rm D}$ in the IVGDR region are larger than those from the $(p,p^\prime)$ data as illustrated in Fig.~\ref{fig:alphad-gdr} and from the particularly large photoabsorption strengths of Ref.~\cite{Fultz1969} in the energy region \unit[$20-30$]{MeV} (cf.~Fig.~\ref{fig:alphad-highenergy}).

The implications of the isotopic dependence and absolute values of the polarizabilities summarized in Tab.~\ref{tab:pol} are discussed in Ref.~\cite{Bassauer2020}. 

\subsection{Magnetic dipole strength}
\label{subsec5c}

The multipole decomposition analysis yields apart of the dominant $E1$ contribution also considerable $M1$ contributions to the cross sections in the PDR region. 
It is possible to determine the IVSM1 strength $B(M1_{\sigma\tau})$ and with some additional assumptions also the corresponding electromagnetic $B(M1)$ strength.
The analysis is based on the so-called unit cross section method and utilizes isospin symmetry of the isovector spin $M1$ mode and the analog Gamow-Teller (GT) mode excited in charge exchange reactions \cite{Fujita2011}. 
In the following only the essential steps of the procedure are sketched. 
A detailed description of the method, the impact of various approximations, and an estimate of systematic uncertainties can be found in Ref.~\cite{Birkhan2016}.

The spin-$M1$ strength is related to the isovector part of the differential $M1$ cross section by
\begin{equation}
	\label{eqn:spinM1}
	\frac{\mathrm{d} \sigma}{\mathrm{d} \Omega}(0^\circ)^{IV}_{exp} = \hat{\sigma}_{M1}F(q, E_x)B(M1_{\sigma\tau}),
\end{equation}
where $\hat{\sigma}_{M1}$ is the unit cross section, $F(q, E_x)$ a kinematic correction factor depending on momentum transfer $q$ and excitation energy $E_{\rm x}$, and $B(M1_{\sigma\tau})$ the dimensionless isovector spin $M1$ strength (the analog of the GT strength for $T = T_0$, where $T_0$ denotes the g.s.\ isospin).

Because of the properties of the effective proton-nucleus interaction \cite{Love1981} at small momentum transfers, for the inelastic proton scattering experiment discussed in this work the spin $M1$ cross sections are predominantly of isovector nature. 
Isoscalar contributions are expected at the level of a few percent and are neglected here. 
Utilizing isospin symmetry, the unit cross section can be taken from analog studies of Gamow-Teller transitions in charge exchange experiments \cite{Taddeuci1987,Zegers2007}. The systematics of the GT unit cross section for $(p,n)$ reactions at $E_p\cong\unit[300]{MeV}$ was investigated in Ref.~\cite{Sasano2009}, where a mass dependent formula for the unit cross section (in mb/sr) was derived
\begin{equation}
	\hat{\sigma}_{GT}=3.4(2)\exp{[-0.40(5)(A^{1/3}-90^{1/3})]}.
\end{equation}
The kinematical correction factor was determined by DWBA calculations and an extrapolation from experimental data at finite angles to the cross section at $0^\circ$ with the aid of the theoretical $M1$ angular distribution shown in Fig.~\ref{fig:120SnCurves}.
Finally, the corresponding electromagnetic strength can be calculated neglecting isoscalar and orbital parts of the electromagnetic $M1$ operator
\begin{equation}
	B(M1) = \frac{3}{4\pi}(g_s^{IV})^2B(M1_{\sigma\tau})\,\mathrm{\mu_N^2},
\end{equation}
where $g_s^{IV}=\frac{1}{2}(g_s^\pi-g_s^\nu)$ is the isovector gyromagnetic factor with proton and neutron $g$ factors $g_s^\pi=5.586$ and $g_s^\nu~=~-3.826$, respectively.

\begin{figure}
	\centering
	\includegraphics[width=\columnwidth]{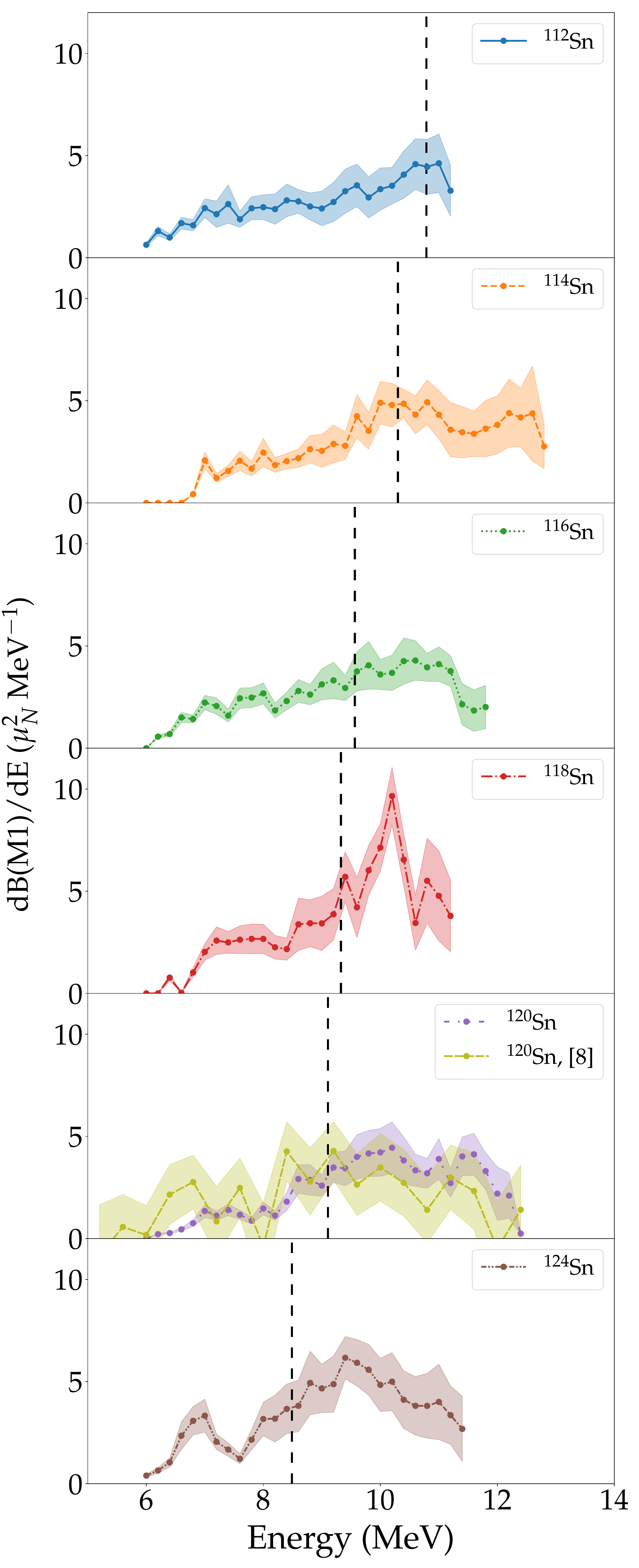}
	\caption{$B(M1)$ strength distributions extracted with the method described in the text. 
Additionally, for $^{120}$Sn results based on the measurement of spin-transfer observables from Ref.~\cite{Hashimoto2015} are shown. 
The vertical lines indicate the neutron threshold energies.}
	\label{fig:bm1}
\end{figure}

Figure \ref{fig:bm1} presents the $B(M1)$ strength distributions applying the above described method to the $M1$ cross sections resulting from the MDA (Fig.~\ref{fig:MDAResults}).
Maximum strength is found between \unit[9]{MeV} and \unit[10.5]{MeV} but the distributions are generally broad, similar to what was observed in heavy deformed nuclei \cite{Heyde2010}.

While the strengths are confined for most isotopes below \unit[11.5]{MeV}, there are two distinct cases where they extend to \unit[12.8]{MeV} ($^{114}$Sn) and \unit[12.4]{MeV} ($^{120}$Sn), respectively.
We note, however, that the MDA results for the $M1$ part of the cross sections has potentially large systematic uncertainties above the $S_n$ energies (indicated by vertical lines in Fig.~\ref{fig:bm1}) due to the similarity of the theoretical $M1$ (Fig.~\ref{fig:120SnCurves}) and the continuum background (Eq.~\ref{eq:qfs}) angular distributions.
While the variation of the sum of both components in the MDA fits is limited (i.e., the $E1$ part is hardly affected), the $M1$ contributions were found to vary strongly in combination with different theoretical $E1$ curves for comparable $\chi^2$ values.
Thus, the $\chi^2$-weighted averaging over the different fits performed in Eq.~(\ref{eq:chi2red}) becomes questionable and Eq.~(\ref{eq:chi2rederror}) underestimates the uncertainties. 
Accordingly, the $B(M1)$ strengths  above the respective $S_n$ values should be taken with some care and additional weak contributions at even higher excitation energies cannot be excluded.
Table \ref{tab:m1} summarizes the results.
\begin{table}
	\centering
	\caption{Neutron threshold energies $S_n$, $B(M1)$ strengths up to $S_n$, and total $B(M1)$ strengths up to energy $E_{\rm max}$ in $^{112,114,116,118,120,124}$Sn deduced from the present data as described in the text.}
	\begin{tabular}{ccccc}
 		\hline
 		\hline
 		& $S_n$ & $\sum_6^{S_n} B(M1)$  & $E_{\rm max}$ &  $\sum_6^{E_{\rm max}} B(M1)$  \\
 		& (MeV) & ($\mu_N^2$) & (MeV) & ($\mu_N^2$)  \\
 		\hline
 		\noalign{\vskip 0.5mm}
 		$^{112}$Sn & 10.79 & 13.1(1.2) & 11.2 & 14.7(1.4) \\
 		$^{114}$Sn & 10.30 & 9.2(1.0) & 12.8 & 19.6(1.9) \\
 		$^{116}$Sn & 9.56  & 8.1(0.7) & 11.8 & 15.6(1.3) \\
 		$^{118}$Sn & 9.32  & 8.2(1.1) & 11.2 & 18.4(2.4) \\
 		$^{120}$Sn & 9.10  & 4.8(0.5) & 12.4 & 15.4(1.4) \\
 		$^{124}$Sn & 8.49  & 5.6(0.6) & 11.4 & 19.1(1.7)  \\
     	\hline
    	\hline
	\end{tabular}
	\label{tab:m1}
\end{table}

It is instructive to compare the $B(M1)$ strength distribution deduced for $^{120}$Sn with results from an independent decomposition of $E1$ and $M1$ cross sections based on the measurement of spin-transfer observables \cite{Hashimoto2015}. 
These results are included in Fig.~\ref{fig:bm1} as olive band. 
Below threshold the results agree within error bars except for the energy region between 6 and \unit[7]{MeV} where Ref.~\cite{Hashimoto2015} finds larger values.
Above threshold one has to take into account that the measured spin-flip probability may contain contributions from quasifree scattering \cite{Baker1997}. 
Thus, the $M1$ cross sections may be overestimated.
%

\section{Conclusions and outlook}
\label{sec6}

In this work the electric and magnetic dipole response of the even-even stable tin isotopes $^{112,114,116,118,120,124}$Sn was extracted in the excitation energy range \unit[$6-20$]{MeV} from  inelastic proton scattering experiments at \unit[295]{MeV} and very forward angles $0^\circ -  6^\circ$.
The individual contributions of different multipoles to the double differential cross sections were extracted by means of an MDA.

Utilizing the virtual photon method, photoabsorption cross sections were extracted from the $E1$ cross section parts.
The results are compared to previous $(\gamma,xn)$ experiments \cite{Fultz1969,Lepretre1974,Sorokin1974,Sorokin1975} and significant differences are observed on the low-energy flank of the IVGDR, particularly pronounced near the neutron threshold, while recent measurements of the $(\gamma,n)$ reaction \cite{Utsunomiya2009,Utsunomiya2011} show good agreement.
Lorentzian fits in the IVGDR energy region show a smooth centroid energy dependence as a function of $A$ consistent with phenomenological models and a constant width. 
A systematic study of the dependence of IVGDR energies on bulk matter properties with an EDF tuned to describe the giant resonances in $^{208}$Pb reveals that the mass dependence is not yet fully reproduced by present-day models, similar to what was concluded for the polarizability \cite{Bassauer2020}.  

The $B(E1)$ strength distributions were determined and compared below the neutron threshold to $(\gamma,\gamma^\prime)$ experiments, where data on $^{112,116,120,124}$Sn are available. Considerably more strength was found for all cases in the present work, confirming previous findings for $^{120}$Sn~\cite{Krumbholz2015}. 
Furthermore, an accumulation of strength has been detected between 6 and \unit[7]{MeV} in all tin isotopes being most prominent in $^{124}$Sn. 
Comparison with results from isoscalar probes for $^{124}$Sn demonstrates that these transitions are of dominant neutron character as expected for the PDR.
At higher excitation energies the $E1$ strength is of pure isovector character.
The differences between the $E1$ strengths deduced from $(p,p^\prime)$ and $(\gamma,\gamma^\prime)$ data indicate 
the influence of complex wave functions of the excited states resulting in small branching ratios to the ground state.

The evolution of the dipole polarizability in the chain of stable tin isotopes was determined combining the experimental photoabsorption cross sections  up to \unit[20]{MeV} from the present work with a theory-aided correction for the unobserved high-energy part.
The implications of these results for the development of EDFs aiming at a global description of the dipole polarizability across the nuclear chart and the resulting constraints on symmetry energy parameters have been discussed in Ref.~\cite{Bassauer2020}. 

Using the unit cross section technique \cite{Birkhan2016}, $B(M1)$ strength distributions were determined from the $M1$ cross sections and a survey of the IV spin $M1$ strength is provided for the first time for stable even-even tin isotopes $^{112-120,124}$Sn.
Below $S_n$ they exhibit broad distributions similar to what was found in heavy deformed nuclei \cite{Heyde2010}.
Above $S_n$, the accuracy is limited because of the similarity of the $M1$ and phenomenological continuum angular distributions in the MDA.

With the $B(E1)$ and $B(M1)$ strength distributions at hand, the Gamma Strength Function (GSF) can be determined for the nuclei studied.
Below neutron threshold, the GSFs show a specific evolution with mass number. 
In combination with compound nucleus $\gamma$-decay experiments using  the Oslo method \cite{Larsen2017} this can provide a unique test of the controversially discussed Brink-Axel hypothesis \cite{Netterdon2015,Bassauer2016,Guttormsen2016,Martin2017,Isaak2019,Fanto2020} stating an independence of the GSF from initial and final states.
Such an analysis is presently prepared \cite{Markova2020}.

Finally, an aspect of the experimental results not discussed here is their high energy resolution of $30-40$ keV (FWHM), which allows a quantitative analysis of the fine structure of the IVGDR similar to Refs.~\cite{Poltoratska2014,Fearick2018}.
Utilizing wavelet analysis techniques \cite{Shevchenko2008}, information on the relevance of different mechanisms contributing to the width of the IVGDR can be retrieved \cite{vonNeumann-Cosel2019a}. 
The cross section fine structure also permits an extraction of the $J^\pi = 1^-$ level density in the IVGDR energy region \cite{Poltoratska2014,Bassauer2016,Martin2017} based on a fluctuation analysis \cite{Kalmykov2006,Kalmykov2007}.
However, this requires excellent statistics which were only reached in the present data for $^{120}$Sn and $^{124}$Sn.
The results will be presented elsewhere \cite{Markova2020}.

\begin{acknowledgements}

The experiments were performed at RCNP under program E422. 
The authors thank the accelerator group for providing excellent beams.
This work was funded by the Deutsche Forschungsgemeinschaft (DFG, German Research Foundation) under grant No.\ SFB 1245 (project ID 279384907), by JSPS KAKENHI (grant No.\ JP14740154), and by MEXT KAKENHI (grant No.\ JP25105509).
C.A.B.\ was supported in part by U.S.\ DOE Grant No.\ DE-FG02-08ER41533 and U.S.\ NSF Grant No.\ 1415656.

\end{acknowledgements}

\bibliographystyle{apsrev4-2} 
\bibliography{Literatur }

\end{document}